\documentclass[a4paper,11pt]{article}
\pdfoutput=1 

\usepackage{jheppub} 

\usepackage[T1]{fontenc} 

\title{\boldmath F-theory Family Unification}

\author[a
]{S. Mizoguchi
}

\affiliation[a]{Theory Center, Institute of Particle and Nuclear Studies,
KEK, \\Tsukuba, Ibaraki, 305-0801, Japan }

\emailAdd{mizoguch@post.kek.jp}

\abstract{

We propose a new geometric mechanism for naturally realizing 
unparallel three families of flavors in string theory, using the framework 
of F-theory.
We consider a 
set of 
coalesced local 7-branes of a particular Kodaira 
singularity type and allow some of the 
branes 
to bend and separate from the rest, 
so that they meet only at an intersection point. 
Such a local configuration
can preserve supersymmetry. Its matter spectrum is investigated 
by studying string junctions near the intersection, and 
shown  
to coincide,   
after an orbifold projection,  
with 
that of a  
supersymmetric coset sigma model 
whose target space is a homogeneous K\"{a}hler manifold 
associated with a corresponding painted Dynkin diagram.
In particular, 
if one starts from the $E_7$ singularity, one obtains 
the $E_7/(SU(5)\times U(1)^3)$
model 
yielding  
precisely 
three generations 
with an unparallel family structure. 
Possible applications to string phenomenology are
also discussed.

}

\font\mybb=msbm10 at 12pt

\font\myeu=eufm10 at 12pt
\def\bb#1{\hbox{\mybb#1}}

\def\ZZ {\bb{Z}}

\def\CCC {\bb{C}}
\def\PP {\bb{P}}
\def\HH {\bb{H}}
\def\II {\bb{I}}

\def\AA{{\bf A}}
\def\BB{{\bf B}}
\def\CC{{\bf C}}

\def\aaa{{\bf a}}

\def\ccc{{\bf c}}
\def\xxx{{\bf x}}

\newcommand\beqa{\begin{eqnarray}}
\newcommand\eeqa{\end{eqnarray}}
\newcommand\n{\nonumber\\}

\begin{document} 
\maketitle
\flushbottom



\section{Introduction}
%
The LHC experiments finally 
found \cite{LHC1,LHC2} the long sought-for elementary particle that was able 
to complete the Standard Model, the Higgs boson. They have also shown that 
the scale of any new physics beyond the Standard Model must be 
pretty much higher than the electro-weak scale. We are now 
more and more seriously interested in why the Standard Model is as 
it is: Why is the top  quark so heavy? Why is the lepton-flavor mixing 
so large? And in the first place, why are there three generations of 
quarks and leptons?



In the past 30 years after the discovery of superstring theories, 
an enormous amount of knowledge on 
elementary particles has been accumulated.
Requirements, or expectations, for realistic string-phenomenology models
have become more and more demanding.
%
Indeed, the top quark was finally found \cite{CDF,D0} 
in 1995 at Tevatron, and the mass
turned out to be about $10^5$ times heavier than the up quark. 
In 1998, the zenith angle dependence of the muon atmospheric neutrino 
was discovered at SuperKamiokande \cite{SuperKamiokande}, 
where the $\theta_{23}$
lepton flavor mixing angle was revealed to be almost maximal, 
$\sim 45^\circ$.
Later neutrino experiments \cite{KamLAND,SNO} also confirmed that another mixing angle 
$\theta_{12}$ was large, and $\theta_{13}$ was also nonzero \cite{T2K,DoubleChooz,DayaBay,RENO}---all these 
undeniable experimental data point to a single fact: {\em The three 
flavors are not on equal footing.}

Superstring theory, however, has developed almost independently of 
these experimental discoveries. 
To say the least, even though 
it could contrive to achieve 
such hierarchical structures (which should be different between the quark 
and lepton sectors, and also between the up and down types) 
by more or less ad-hoc assumptions and/or fine tunings, it has never 
been able to explain them.
Of course, it would be easy to close our eyes to 
all these facts and dismiss everything as
an accident; this is not our attitude in this paper.

There have been numerous efforts to understand the hierarchical family 
structure of quarks and leptons. Particularly interesting among them is 
the idea of {\em family unification}.  Family unification is the idea that 
the quarks and leptons are the fermionic partners of the scalars of 
some {\em coset} supersymmetric nonlinear sigma model 
\cite{BPY,BLPY,OngPRD27,Ong,KY,BKMU,IY,Buchmuller:1985rc,IKK}.
%
%
A remarkable observation made 
by Kugo and Yanagida was that \cite{KY} 
the supersymmetric sigma model based on 
$E_7/(SU(5)\times SU(3)\times U(1))$ 
had precisely {\em three} sets of 
${\bf 10}\oplus {\bf\bar 5}$ of $SU(5)$, 
%
%
in addition to a single ${\bf 5}$\footnote{and hence it is anomalous 
both as a gauge theory and as a nonlinear sigma model.  
It was pointed out \cite{YanagidaYasui} that these anomalies were removed by introducing 
matter chiral-multiplets. This issue is further discussed in 
section \ref{Orbifolds}.}, 
as its target space. 
What is special here is that 
 the three generations 
are asymmetrically embedded  into $E_7$ as $SU(5)$ multiplets.
Indeed, the two ${\bf\bar 5}$'s, 
identified as the second and third generations of the $SU(5)$ GUT 
multiplets \cite{SUSYGUT,SakaiSUSGUT} containing the down-type quarks, 
come with the ``symmetry breaking'' from $E_7$ to $E_6$, whereas 
the last ${\bf\bar 5}$ arises when $E_6$ ``breaks'' to $SO(10)$. 
In contrast, a ${\bf 10}$ representation arises at each step when 
the rank of the ``symmetry group'' is reduced by one 
\begin{figure}[t]%
\centering
\includegraphics[height=0.25\textheight]{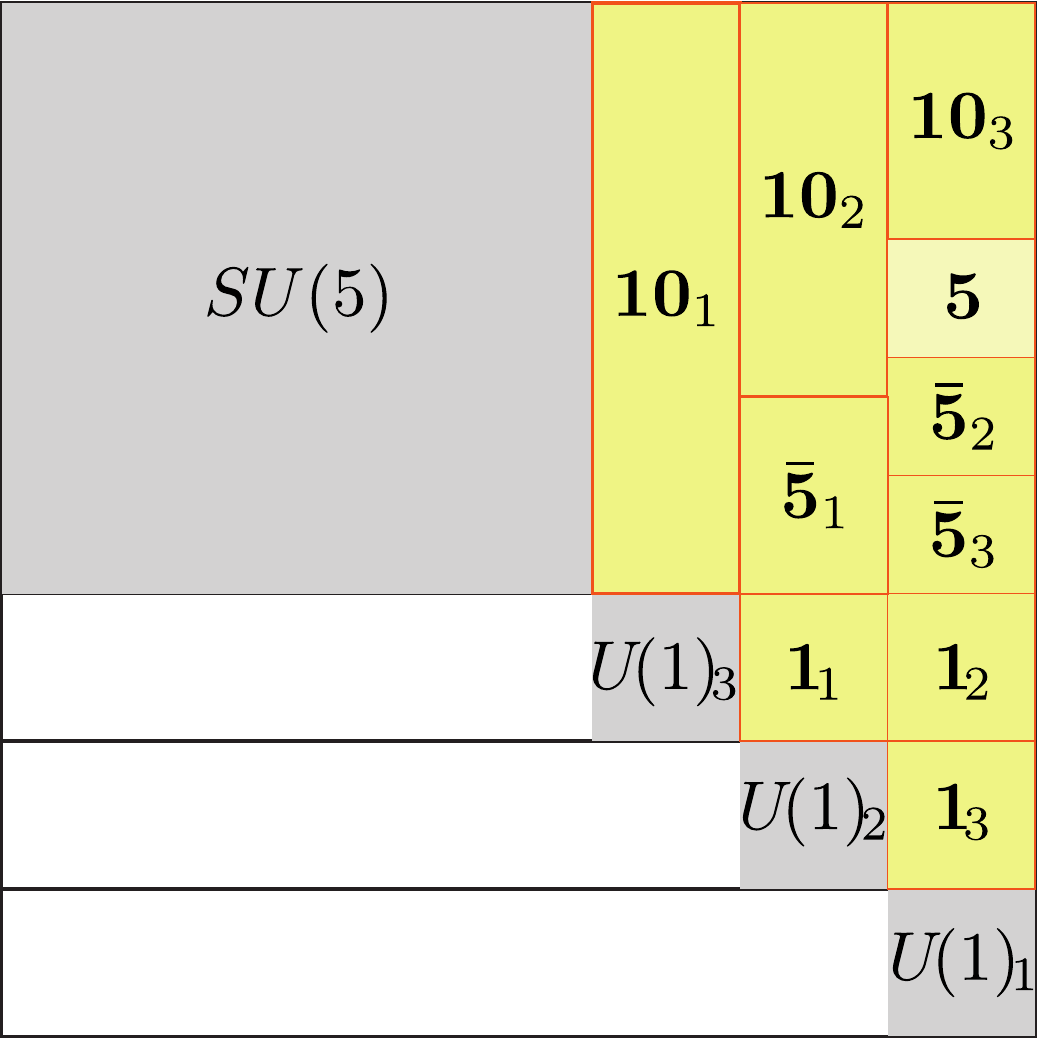}
\caption{\label{KugoYanagida_coset}  The $E_7/(SU(5)\times U(1)^3)$ model. 
The next larger square than the shaded ``$SU(5)$'' represents $SO(10)$, 
the next to next is $E_6$, and the whole square is $E_7$.
Exactly three sets of $SU(5)$ multiplets for quarks and leptons 
perfectly fit (except one {\bf 5}) in the coset. 
The $E_7/(SU(5)\times SU(3) \times U(1))$ model consists of 
only the non-singlets.}
\end{figure}
(Figure \ref{KugoYanagida_coset})
\footnote{
If one considers $E_7/(SU(5)\times U(1)^3)$ instead, then
one has (with an appropriate choice of the complex structure) 
three sets of ${\bf 10}\oplus {\bf\bar 5} \oplus {\bf 1}$
and one ${\bf 5}$. The fact that the coset 
$E_7/SU(5)$ accommodates three generations of quarks and leptons 
was already noted in \cite{BPY}, but the $U(1)$ factors (relevant to 
the K\"{a}hler structure) in the 
denominator group were not specified. The $E_7$ coset model 
was also mentioned in \cite{OngPRD27}. The importance 
of the coset $E_7/(SU(5)\times U(1)^3)$ as well as 
$E_7/(SU(5)\times SU(3) \times U(1))$ was emphasized in 
\cite{YanagidaYasui}, where the issue of anomaly cancellation in these models 
was also addressed. In this paper, we also call 
the $E_7/(SU(5)\times U(1)^3)$ 
(as well as 
$E_7/(SU(5)\times SU(3) \times U(1)$) model {\em the Kugo-Yanagida model}
as it is obvious that the former differs from the latter only by some singlets.}. 
This was called the ``unparallel'' family structure by Yanagida \cite{Yanagidatalk,SatoYanagida}.


Therefore, a natural question to ask is: Can one get 
such a coset 
family structure in string theory? 
%
%
So far, very few attempts have been made to obtain Kugo-Yanagida-type
models in string theory, and the few have met with 
only partial success \cite{KM,MY}.

In this paper, we take a modest 
step towards  
realizing this unparallel family structure in string theory. 
We will show that such a 
structure can in fact naturally arise in string theory, 
in the framework of local F-theory.
The study of phenomenological applications of F-theory \cite{EvidenceforFtheory,Ftheoryorientifold} has been 
of much interest in the past several years 
\cite{DonagiWijnholt,Fmodel24,Fmodel23,Fmodel22,Fmodel21,Fmodel20,Fmodel19,Wijnholt,
Fmodel18,Fmodel17,Fmodel16,Fmodel15,Fmodel14,Fmodel13,
Fmodel12,Fmodel11,Fmodel10,Fmodel9,Fmodel8,Fmodel7,Fmodel6,Choi:2009tp,
Fmodel5,Fmodel4,Fmodel3,CGEH_Global_Ftheory,Fmodel2,Fmodel1,ChoiKyae,Choi}, 
but in this paper 
we will bring a slightly different perspective. 
We show that a certain local 7-brane system in F-theory 
can realize, already at the level of six dimensions, the same 
quantum numbers as that of the SUSY nonlinear sigma model 
considered in family unification.
Our key observation is that the representations of the 
charged matter hypermultiplets arising at the singularity 
are precisely (as far as the enhanced singularity is of 
the split type \cite{BIKMSV}) the ones consisting of a 
homogeneous K\"ahler manifold of the corresponding painted Dynkin diagram.
In particular, 
if one starts from the $E_7$ singularity, one obtains a set of {\em six-}dimensional 
massless matter which have 
the same quantum numbers as those of
the $E_7/(SU(5)\times U(1)^3)$
Kugo-Yanagida model. 
%

%
%
We should note that our mechanism itself does not yet 
ensure three generations 
of ${\bf 10}\oplus {\bf\bar 5}$ in a {\em four}-dimensional F-theory compactification 
as it is still a six-dimensional analysis.
The most modern and common way of achieving a chiral spectrum is
to turn on so-called G-fluxes \cite{DonagiWijnholt}.
Alternatively, however, a practical approach to get a four-dimensional 
chiral theory is to compactify two of the the six dimensions on a 
two-torus and project out half of the spectrum by taking an orbifold.
This is discussed in section \ref{Orbifolds}.
We also note that the requisite quantum numbers are already 
obtained in six dimensions, and there is no need for contrived 
assumptions to get the desired spectrum here.

Although these rules themselves must have been 
known for some time, the relation 
to homogeneous K\"{a}hler spaces or nonlinear sigma models seems to have 
never been discussed in the literature
\footnote{The coset structure of chiral matter was already  
implied in \cite{DonagiWijnholt,Wijnholt}, but the relevance of
string junctions to {\em matter} generation or the possible application 
to family unification was not discussed. The relevance of string junctions 
for the chiral matter generation was first emphasized in \cite{Tani}, 
and also more recently in \cite{CGEH_Three_looks}.}.
Indeed, as of writing this article, 
there is only one paper \cite{TatarWatari} in the INSPIRE database that cites 
both Katz-Vafa \cite{KatzVafa} and Kugo-Yanagida \cite{KY},
and in \cite{TatarWatari} such a connection was not mentioned.

We are interested in some {\em local} geometric structure 
that can realize precisely three unparallel families. This is 
because if the realization of the SM were a consequence of 
the global details of the entire compactification space, it would be 
very hard, if not impossible, to find any reason or explanation  
for what we observe now. 

The plan of the rest of this paper is as follows: In section \ref{CosetSpaceFamilyUnification}, 
we give a brief review of the basic idea of coset space family unification.
In section \ref{section3}, we first recall the known results of 
F-theory/heterotic duality in six dimensions, and then explain 
Tani's argument of how the chiral matter at the extra zeroes of 
the discriminant can be understood in terms of string junctions. 
We are naturally led to the coset structure of the chiral matter, and 
propose the 7-brane configuration for the $E_7/(SU(5)\times U(1)^3)$ 
model. In section \ref{SUSYdeformation}, we prove that the 7-brane configuration can 
preserve SUSY.  Section  \ref{Orbifolds} is devoted to a brief discussion on 
how we can derive a four-dimensional model from the six-dimensional 
one obtained from the configuration. It is also pointed out that 
in our setup there is a possibility for the anomalies of the original model 
to cancel due to the anomaly inflows. 
In section \ref{ExplicitCurve},  we present the explicit local expression for the 
curve of the brane configuration. Finally, we conclude in section  \ref{Discussion} 
with a summary and discussion. Appendix \ref{appendixA}  contains the explicit 
result of the recursion relation in section \ref{SUSYdeformation}, whereas Appendix \ref{appendixB}
is a brief explanation of how the monodromy is read off by tracing 
the value of the $J$ function.

%
%

\section{Coset space family unification}
\label{CosetSpaceFamilyUnification}
We begin with a review of the basic idea of coset space family unification, which 
is what we want to achieve in string theory.
We will be brief, and for more detailed discussion we refer the reader to 
\cite{IKK}, and also \cite{BFR}. 

As we already mentioned in Introduction, family unification is the 
idea that 
the quarks and leptons 
can be understood  
as quasi-Nambu-Goldstone 
fermions \cite{BLPY}
of a supersymmetric {\em coset} 
nonlinear sigma model \cite{BPY,Ong,KY,IY,BKMU,IKK}.
This means that the target space of the sigma model is some 
homogeneous space $G/H$ associated with a  
Lie group $G$ and its closed subgroup $H$.
The idea of identifying 
all the three families as being 
a part of some group
is an old one \cite{E63,oldFU1,oldFU2,oldFU3,oldFU4,oldFU5,oldFU6,
oldFU7,oldFU8,oldFU9,oldFU10,oldFU11}, 
going back before the superstring theories were found,
but 
it is important to 
note that being a coset is essential for the chiral nature of the 
spectrum, which is in contrast to the models in the earlier literature.

Originally, such a nonlinear sigma model was thought of as 
arising from a spontaneous supersymmetry breaking 
of some global symmetry caused by a strong gauge dynamics of 
the underlying ``preon'' theory \cite{BPY}.
Later, we will show an alternative, geometric origin of these 
sigma models in F-theory.

To characterize $D=4$, ${\cal N}=1$ supersymmetric sigma models 
the following two facts are essential: The first fact is that, in order to 
have $D=4$, ${\cal N}=1$ SUSY, the scalar manifold must be 
K\"{a}hler \cite{Zumino, AGF}, which is well-known. The second is 
a classic result due to Borel \cite{Borel}: Let $G$ and $H$ be a semi-simple 
Lie group and its closed subgroup,
then the coset space $G/H$ is K\"{a}hler if and only if $H$ 
is the group consisting of all elements that commute with  
some $U(1)^n$ subgroup of $G$. Thus it follows that 
every element in $G/H$ has a nonzero charge for some 
$U(1)$ subgroup, since otherwise such an element must belong 
to $H$ by construction. 
Borel's theorem also states that the set of all $G$-invariant complex structures 
correspond one-to-one to that of all Weyl chambers of the Lie 
algebra. This statement can be translated into a useful 
way of distinguishing different complex structures as follows \cite{IKK}: 
Suppose that we have chosen some $U(1)$ generator $Y$
%
%
\footnote{Note that, despite its notation, 
``$Y$-charge'' here is {\em not} the weak hypercharge 
of the SM. We use this term following \cite{IKK}.}
which commutes with any generator of $H$. We can 
decompose the Lie algebra of $G$ into a direct sum of eigenspaces 
of ${\rm ad} Y$, that is, into a sum of spaces of ``states'' with different 
$U(1)$ ``$Y$-charges''. Then $G/H$ consists of the spaces with 
negative $Y$-charges. If the charge vector of $Y$ is 
varied so that it passes across into the next Weyl chamber, then 
one of the signs of the $Y$-charges flips, and this corresponds to 
the change of the complex structure.

Let us illustrate the above with an example, which is the main 
focus of the subsequent discussion.
The Lie algebra $E_7$ is decomposed into a sum of 
irreducible representations of $SU(5) \times SU(3)$ as:
\beqa
{\bf 133}&=&
({\bf 24},{\bf 1})_0\oplus
({\bf 1},{\bf 8})_0\oplus
({\bf 1},{\bf 1})_0\oplus
({\bf 5},{\bf \bar 3})_{4}\oplus
({\bf \bar 5},{\bf 3})_{-4}
\n
&&\oplus 
({\bf 5},{\bf 1})_{-6}\oplus
({\bf \bar 5},{\bf 1})_{6}\oplus
({\bf 10},{\bf \bar 3})_{-2}\oplus
({\bf \overline{10}},{\bf 3})_{2}.
\label{E_7decomposition}
\eeqa
The $U(1)$ subgroup that commutes with $SU(5) \times SU(3)$ 
is uniquely determined, and its charges are indicated 
as subscripts. 
Collecting only the representations that have negative charges, 
we find that the homogeneous space 
$E_7/(SU(5)\times SU(3) \times U(1))$ consists of
\beqa
({\bf \bar 5},{\bf 3})_{-4}\oplus({\bf 10},{\bf \bar 3})_{-2} \oplus ({\bf 5},{\bf 1})_{-6}
\label{KYspectrum}
\eeqa
as advocated.
On the other hand, for $E_7/(SU(5)\times U(1)^3)$ the $SU(3)$ multiplets 
in (\ref{KYspectrum}) are further decomposed, and besides, three more 
singlets emerges from the $SU(3)/U(1)^2$ piece.
Their $U(1)$ charges are summarized in Table \ref{U(1)charges} \cite{SatoYanagida},
where the three $U(1)$'s are such that 
$E_7 \supset E_6\times U(1)_1$, $E_6\supset SO(10)\times U(1)_2$ and 
$SO(10) \supset SU(5) \times U(1)_3$.
\begin{table}
\centering
%
\caption{$U(1)$ charges of the $SU(5)$ multiplets in $E_7/(SU(5)\times U(1)^3)$
 \cite{SatoYanagida}. \label{U(1)charges}}
\begin{tabular}{|c|c|c|c|}
\hline 
$SU(5)$ representation&
$U(1)_1$ charge($=Q_1$)&
$U(1)_2$ charge($=Q_2$)&
$U(1)_3$ charge($=Q_3$)\\
\hline
${\bf 10}_1$
&$0$
&$0$
&$4$
\\
${\bf 10}_2$
&$0$
&$3$
&$-1$
\\
${\bf 10}_3$
&$2$
&$-1$
&$-1$
\\
\hline
${\bf \bar 5}_1$
&$0$
&$3$
&$3$
\\
${\bf \bar 5}_2$
&$2$
&$-1$
&$3$
\\
${\bf \bar 5}_3$
&$2$
&$2$
&$-2$
\\
\hline
${\bf 1}_1$
&$0$
&$3$
&$-5$
\\
${\bf 1}_2$
&$2$
&$-1$
&$-5$
\\
${\bf 1}_3$
&$2$
&$-4$
&$0$
\\
\hline
${\bf 5}$
&$2$
&$2$
&$2$
\\
\hline 
\end{tabular}
\end{table}
Let $Q_i$ be the $U(1)_i$ charge for $i=1,2,3$, then the $U(1)$ 
$Y$-charge (determining the complex structure) 
in the previous $E_7/(SU(5)\times SU(3) \times U(1))$
case is given by the linear combination
\beqa
Y_{E_7/(SU(5)\times SU(3) \times U(1))}&=& -\frac16\left(10 Q_1 + 5 Q_2 + 3 Q_3 \right).
\eeqa
In the present $E_7/(SU(5)\times U(1)^3)$ case, the $Y$-charge can be 
a linear combination of the form 
\beqa
Y_{E_7/(SU(5)\times U(1)^3)}&=& s Q_1 + t(2 Q_1 + Q_2) + u (10 Q_1 +5Q_2 + 3 Q_3) 
\eeqa 
for any {\em negative} $s,t$ and $u$.

\begin{figure}[h]%
\centering
\includegraphics[width=0.4\textwidth]{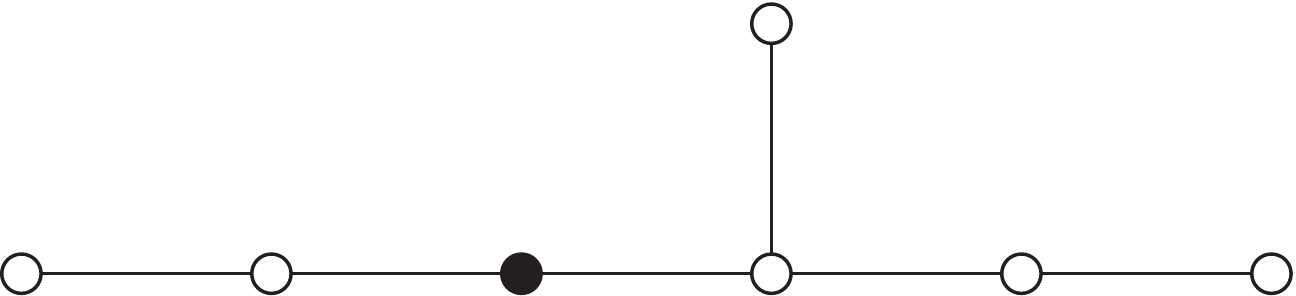}
~~~~~~~~
\includegraphics[width=0.4\textwidth]{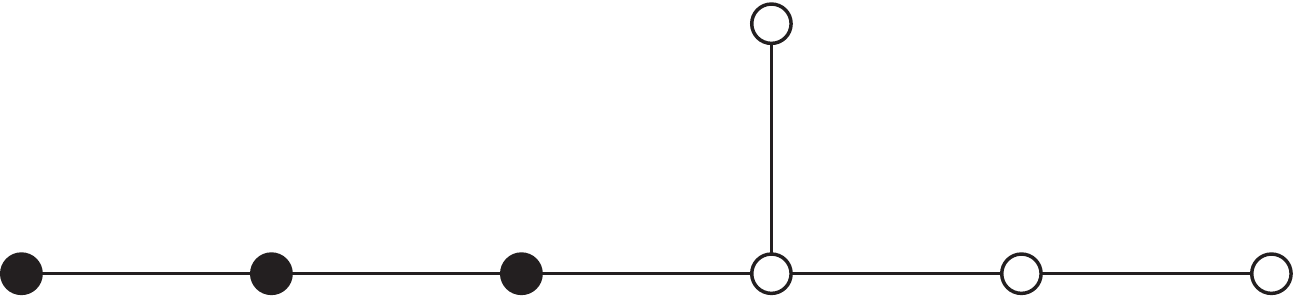}
\caption{\label{paintedE7s}  Painted Dynkin diagrams. Left: $E_7/(SU(5)\times SU(3) \times U(1))$,
right: $E_7/(SU(5)\times U(1)^3)$.}
\end{figure}

As we said, the target space of a coset supersymmetric nonlinear 
sigma model is a homogeneous K\"{a}hler manifold. There is available a 
useful representation of homogeneous K\"{a}hler manifolds in terms of 
painted Dynkin diagrams \cite{BFR}; their ``cook-book recipe'' states that 
\cite{BFR} one first draws the Dynkin diagram for the numerator Lie group, 
paints a subset of the vertices (note that the roles of the black 
(painted) and white nodes are traded here) for the denominator $U(1)$ 
subgroups so that the remaining white Dynkin diagrams correspond to 
the semi-simple part of the denominator group. In this way all 
homogeneous K\"{a}hler manifolds are classified \cite{BFR}. The corresponding 
painted Dynkin diagrams for $E_7/(SU(5)\times SU(3) \times U(1))$
and $E_7/(SU(5)\times U(1)^3)$ are shown in Figure \ref{paintedE7s}.

Basically, this kind of family unification models utilize
the Froggatt-Nielsen mechanism \cite{FNmechanism} to account for
 the origin of the hierarchical family structure.
The $E_7/(SU(5)\times SU(3) \times U(1))$ or 
$E_7/(SU(5)\times U(1)^3)$ model has been investigated by many 
authors from various points of view \cite{E7cosetmodels1,E7cosetmodels2,
E7cosetmodels3,E7cosetmodels4,E7cosetmodels5,E7cosetmodels6,
E7cosetmodels7,E7cosetmodels8,E7cosetmodels9,E7cosetmodels10,
E7cosetmodels11} 
\footnote{Also related is the idea of ``$E_6$ unification'' 
\cite{E61,E62,E63,E64,E65,E66,E67,E68,E69,E610,E611,
E612,E613,E614,E615,E616,E617}, 
where the origin of the difference between the flavor structures 
of quarks and leptons is attributed to the asymmetry between the 
${\bf 10}$ and ${\bf \bar 5}$ representations in a ${\bf 27}$ 
multiplet of $E_6$.}.

\section{F-theory/heterotic duality, string junctions and 
the correspondence to homogeneous K\"{a}hler 
manifolds}
\label{section3}
\subsection{Review of F-theory/heterotic duality in six dimensions}

In this section we recall the beautiful results on F-theory/heterotic duality 
in six dimensions. Although this has already been well known for some time, 
it is useful to review the original discussion because thereby its connection 
to homogeneous K\"{a}hler manifolds can be uncovered.

The proposal of 
\cite{MorrisonVafa} was that the $K3$ compactification of the $E_8\times E_8$
heterotic string with instanton numbers $(12-n, 12+n)$ is dual to F-theory 
compactified on an elliptic Calabi-Yau three-fold over the base space 
%
%
being the Hirzeburch surface ${\bf F}_n$. More precisely, the dual 
Calabi-Yau manifold was given in the Weierstrass form as 
\cite{MorrisonVafa,KachruVafa} 
\beqa
y^2&=&x^3+x\sum_{i=0}^8 z^i f_{8+(4-i)n}(w)+\sum_{j=0}^{12} z^j g_{12+(6-j)n}(w),
\label{dualCY}
\eeqa
where $z$ is the coordinate for the ${\bf P}^1$ fiber of ${\bf F}_n$,
and $w$ is the one for the ${\bf P}^1$ base of ${\bf F}_n$. 
$f_k(w)$, $g_k(w)$ are $k$-th order polynomials of $w$.
Intuitively, this six-dimensional duality is understood as the fiber-wise 
duality in eight dimensions \cite{EvidenceforFtheory} further compactified 
and fibered over ${\bf P}^1$. 

This proposed duality was examined in detail in \cite{BIKMSV}. In 
particular, the dimensions of the neutral hypermultiplet moduli spaces 
were compared between the two, and a perfect match was 
found in various cases of unbroken gauge 
symmetries. It was also found there \cite{BIKMSV} that 
the charged matter arose at ``extra zeroes'' of the discriminant 
of the curve (\ref{dualCY}) at particular values of $w$, where the 
singularities of the coinciding gauge 7-branes were (more) enhanced.

For example, let us consider
\beqa
y^2&=&x^3+ z^3 f_{8+n}(w) x +z^5 g_{12+n}(w).
\label{E7curve}
\eeqa
The discriminant is
\beqa
\Delta&=&108 z^9 (f_{8+n}^3 + g_{12+n}^2 z).
\eeqa
%
%
There is (generically) 
an $E_7$ singularity at $z=0$ since $ord(\Delta)=9$,  
$ord(\mbox{coefficient of }x^1)=3$ and
$ord(\mbox{coefficient of }x^0)=5$. Here  
$ord(\cdots)$ denotes the order of $\cdots$ as a polynomial of $z$.
(\ref{E7curve}) is the curve for the F-theory dual to heterotic on $K3$ 
with unbroken $E_7$ gauge symmetry with $12+n$ instantons 
embedded in $SU(2)$.

On the heterotic side, the number of neutral 
hypermultiplets is $2n+21$
\cite{KachruVafa,GreenSchwarzWest},
whereas on the F-theory side, it is determined by the 
dimensions of the complex structure moduli. The latter is the number 
of coefficients of the polynomials up to an overall rescaling:
$(9+n)+(13+n)-1$, which indeed coincides with the heterotic computation.

On the other hand, charged matter in this heterotic compactification 
is found to be \cite{KachruVafa,GreenSchwarzWest} 
$8+n$ half-hypermultiplets in ${\bf 56}$ of $E_7$. 
Since $8+n$ is the number of zeroes $f_{8+n}(w)$, 
these ``extra'' zeroes in the discriminant implied the 
appearance of charged matter in F-theory. Indeed, 
this was confirmed in \cite{BIKMSV} 
in various patterns of gauge symmetry breaking.
Their results are summarized \footnote{We only consider the split case
\cite{BIKMSV} here.} in the first three columns of Table \ref{table1} (together with 
the corresponding set of coalesced 7-branes and associated 
homogeneous K\"ahler manifolds, which are explained shortly).

\renewcommand{\arraystretch}{0.8}
\begin{table}
\caption{\label{table1}Summary of matter fields in F-theory/heterotic 
duality in six dimensions. Only the cases for the split type with rank$\geq 2$ 
are listed, where $n$ is $\pm$(the number of instantons $-12$) in one of $E_8$'s 
on the heterotic side, and $r$ specifies how they are distributed when 
the commutant group is a direct product \cite{BIKMSV}. 
In addition to the data shown in \cite{BIKMSV}, the corresponding 
7-brane configurations as well as the homogeneous K\"{a}hler manifolds 
are also displayed. 
}%
%
\begin{tabular}{|c|c|c|c|c|}
\hline
Gauge group
& Neutral hypers & Charged matter         & 7-branes        &  
$
\renewcommand{\arraystretch}{0.5}
\begin{array}{c}
 \mbox{Homogeneous} \\   \mbox{K\"ahler manifold}   
\end{array}
$
 \\
\hline
$E_7$ & $2n+21$ & $\frac {n+8}2{\bf 56}$          & \AA + $\AA^6$\BB\CC\CC & $E_8/(E_7\times U(1))$ \\
&&&&\scalebox{0.25}{%
 \includegraphics{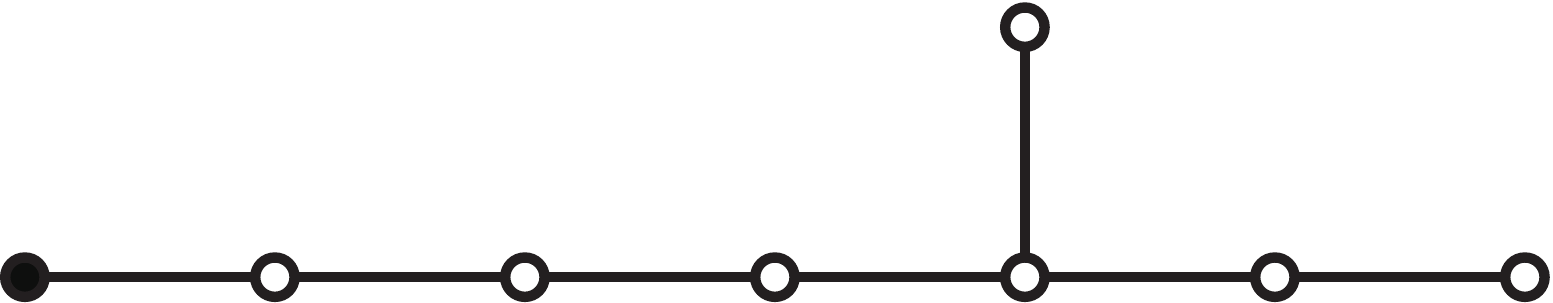}
} \\
\hline
$E_6$ & $3n+28$ & $(n+6){\bf 27}$          &\AA + $\AA^5$\BB\CC\CC & $E_7/(E_6\times U(1))$ \\
&&&&\scalebox{0.25}{%
 \includegraphics{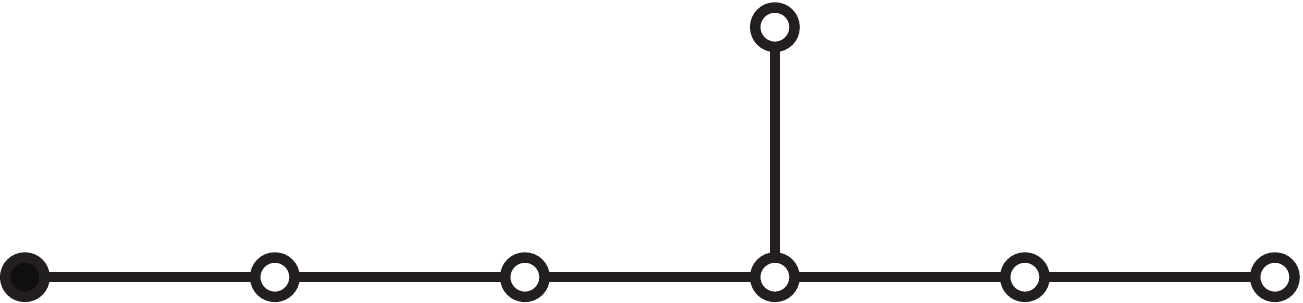}
}\\
\hline
&& 
 $(n+4){\bf 16}$ & $\AA^5$\BB\CC + \CC & $E_6/(SO(10)\times U(1))$\\
$SO(10)$&$4n+33$&&&\scalebox{0.25}{%
 \includegraphics{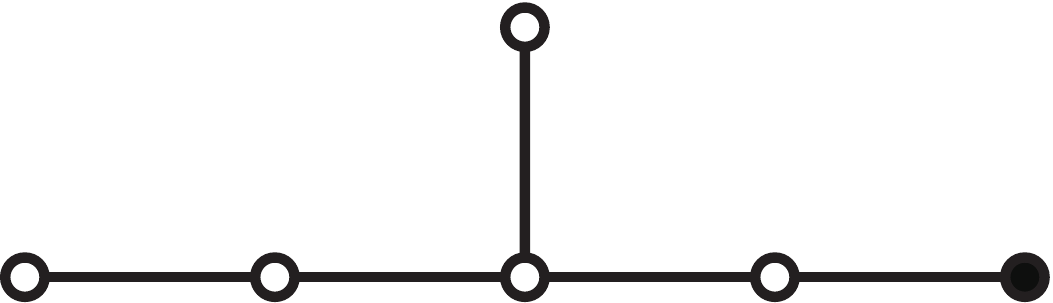}
}\\
		  &			& $(n+6){\bf 10}$ &   \AA + $\AA^5$\BB\CC & $SO(12)/(SO(10)\times U(1))$\\&&&&\scalebox{0.25}{%
 \includegraphics{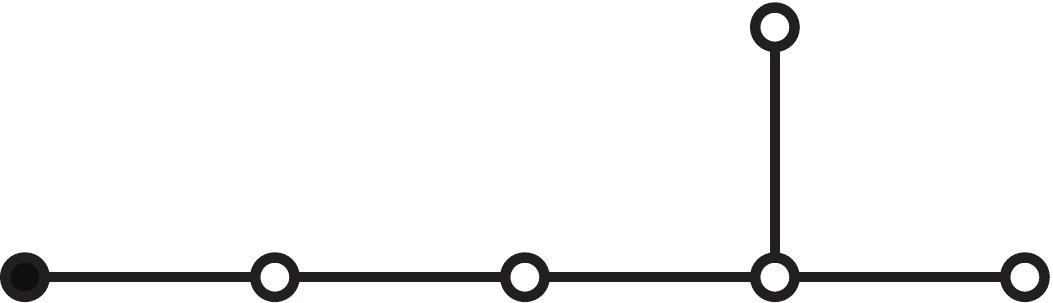}
}\\
\hline
\raisebox{-10pt}{$SO(8)$}& \raisebox{-10pt}{$6n+44$}& 
$
\renewcommand{\arraystretch}{0.5}
\begin{array}{c}
 (n+4){\bf 8}_c \\   (n+4){\bf 8}_s 
\end{array}
$
&$\AA^4$\BB\CC + \CC &$\renewcommand{\arraystretch}{0.5}
						\begin{array}{c} E_5/(SO(8)\times U(1))\\
						 (=SO(10)/(SO(8)\times U(1)))\end{array}$
\\
&&$(n+4){\bf 8}_v$ & \AA + $\AA^4$\BB\CC & $SO(10)/(SO(8)\times U(1))$\\

&&&&\scalebox{0.25}{%
 \includegraphics{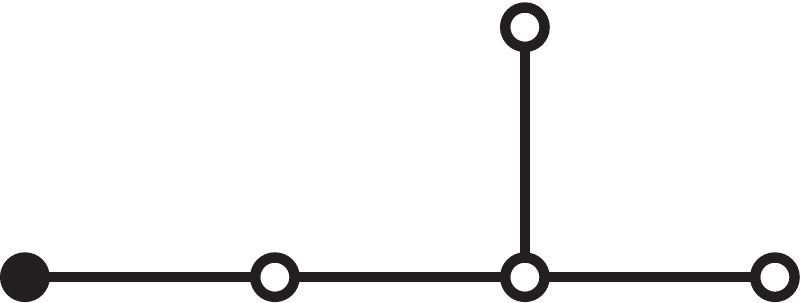}
}
\\
\hline
& & 
      $(4n+16){\bf 4}$ & $\AA^3$\BB\CC + \CC & $\renewcommand{\arraystretch}{0.5}
						\begin{array}{c} E_4/(SO(6)\times U(1))\\
						 (=SU(5)/(SU(4)\times U(1)))\end{array}$
     \\  
 $SU(4)$  &$8n+51$&&&\scalebox{0.25}{%
 \includegraphics{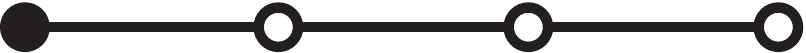}
}\\
     
&&  $(n+2){\bf 6}$    &   \AA + $\AA^3$\BB\CC & $SO(8)/(SO(6)\times U(1))$\\
&&&&\scalebox{0.25}{%
 \includegraphics{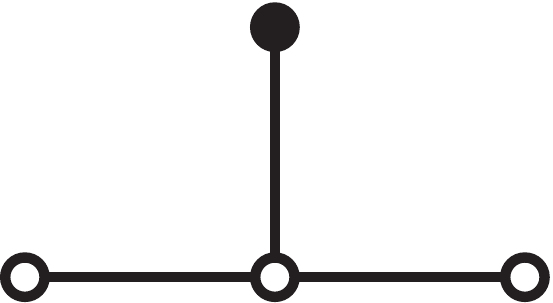}
}\\
\hline
&& 
$
\renewcommand{\arraystretch}{0.5}
\begin{array}{c} (4n+16)(({\bf 1},{\bf 2})
 \\  +({\bf 2},{\bf 1})) 
\end{array}
$
&$\AA^2$\BB\CC + \CC &$\renewcommand{\arraystretch}{0.5}
						\begin{array}{c} E_3/(SO(4)\times U(1))\\
						 (=SU(3)/(SU(2)\times U(1)))\end{array}$\\
$SO(4)$&$10n+54$&&&\scalebox{0.25}{%
 \includegraphics{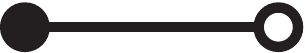}
}\\
&&$n({\bf 2},{\bf 2}) $ & \AA + $\AA^2$\BB\CC & $SO(6)/(SO(4)\times U(1))$\\
 &&&&\scalebox{0.25}{%
 \includegraphics{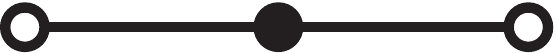}
}\\
\hline
$SU(3)$ & $12n+66$ & $(6n+18){\bf 3}$          &\AA + $\AA^3$ & $SU(4)/(SU(3)\times U(1))$ \\
 &&&&\scalebox{0.25}{%
 \includegraphics{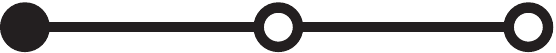}
}\\
%
%
%
\hline
&&
      $\frac r2{\bf 32}+ \frac{n+4-r}2{\bf 32'}$      
      
      & $\AA^6$\BB\CC  +\CC & $E_7/(SO(12)\times U(1))$
     \\
$SO(12)$&$2n+18$&&&\scalebox{0.25}{%
 \includegraphics{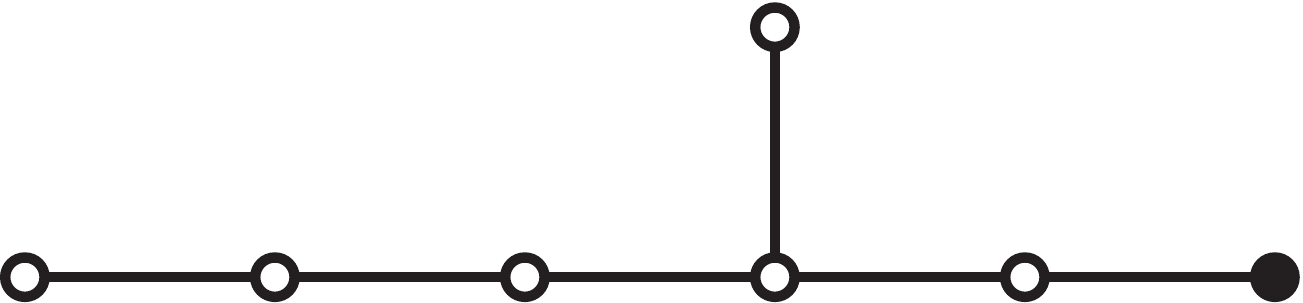}
}\\
&&  $(n+8){\bf 12}$    &   \AA + $\AA^6$\BB\CC & $SO(14)/(SO(12)\times U(1))$\\
     &&&&\scalebox{0.25}{%
 \includegraphics{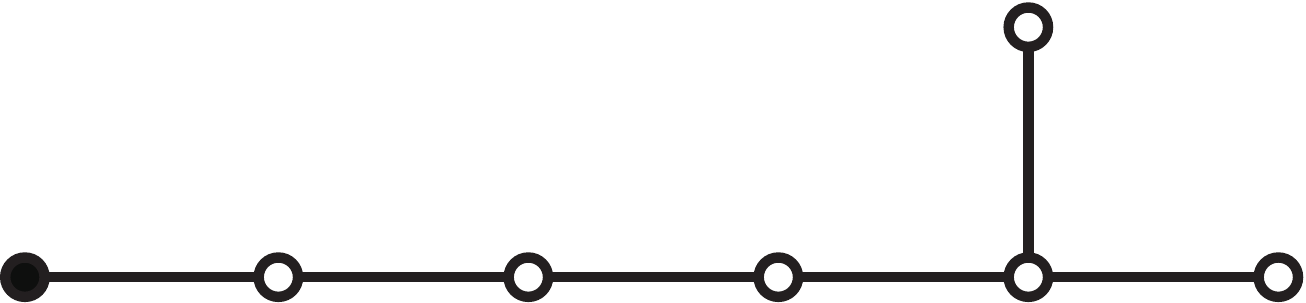}
}\\
\hline
\end{tabular}
\end{table}

\addtocounter{table}{-1}
\begin{table}
\caption{(Cont'd)
}%

\centering
\begin{tabular}{|c|c|c|c|c|}
\hline
& & 
      
      $\frac r2{\bf 20}$

      & $\AA^6$ +${\bf X}_{[2,-1]}$+\CC  
      & $E_6/(SU(6)\times U(1))$\\
     &&&&\scalebox{0.25}{%
 \includegraphics{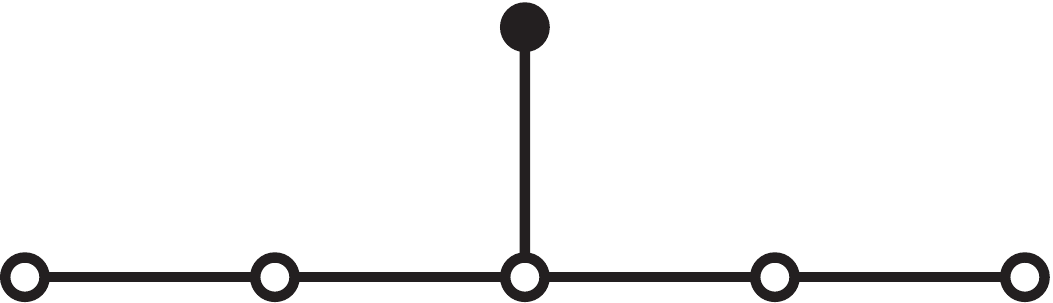}
}\\
$SU(6)$&$3n-r+21$&  

$(2n+16+r){\bf 6}$    &   \AA+ $\AA^6$ & $SU(7)/(SU(6)\times U(1))$\\
     &&&&\scalebox{0.25}{%
 \includegraphics{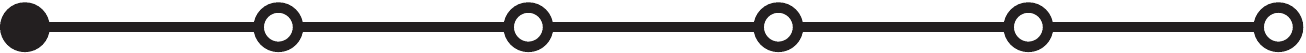}
}\\

& & 
      
      $(n+2-r){\bf 15}$      
      
      & $\AA^6$ +\BB+\CC 
      & $SO(12)/(SU(6)\times U(1))$
     \\
         &&&&\scalebox{0.25}{%
 \includegraphics{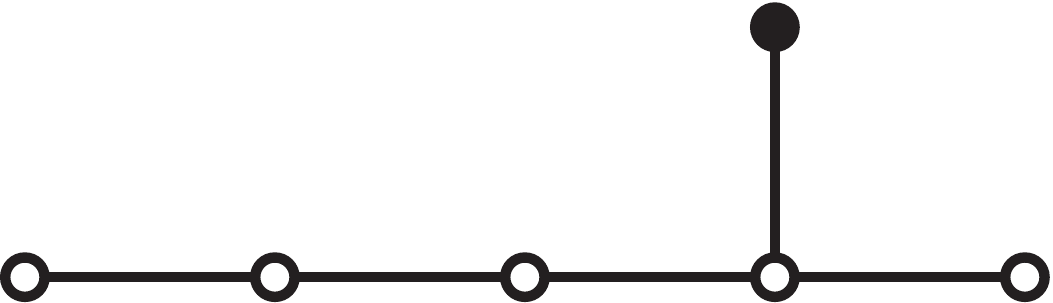}
}\\ 
\hline
& & 
      $(3n+16){\bf 5}$ & \AA+ $\AA^5$ & $SU(6)/(SU(5)\times U(1))$
     \\
        $SU(5)$&$5n+36$&&&\scalebox{0.25}{%
 \includegraphics{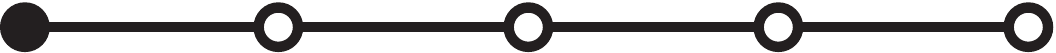}
}\\ 
&&  $(n+2){\bf 10}$    &   $\AA^5$ +\BB+\CC & $SO(10)/(SU(5)\times U(1))$\\
&&&&\scalebox{0.25}{%
 \includegraphics{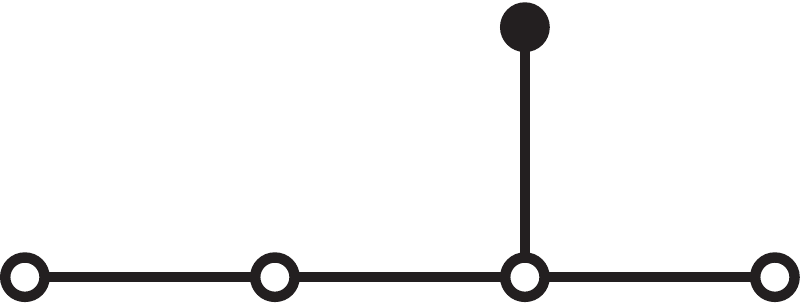}
}\\ 
\hline
\end{tabular}
\end{table}

Note that 
at $8+n$ zero loci of $f_{8+n}(w)$, the discriminant of the first term vanishes, 
becoming a tenth-order polynomial in $z$. Again, 
this means that the singularity is enhanced from 
$E_7$ to $E_8$ at these ``extra zeroes''.

Another example is the dual curve with an $E_6$ unbroken gauge symmetry:
\beqa
y^2&=&x^3+ z^3 f_{8+n}(w) x +z^4 g_{12+2n}(w)+z^5 g_{12+n}(w).
\label{E6curve}
\eeqa
The difference from (\ref{E7curve}) is that it contains a $z^4$ term, and also 
$g_{12+2n}(w)$ must be in the split form \cite{BIKMSV}, that is 
\beqa
g_{12+2n}(w)&=q_{6+n}(w)^2
\eeqa
for some $q_{6+n}(w)$. Then 
the discriminant becomes
\beqa
\Delta&=&27 z^8 q_{6+n}^4+ z^9(4f_{8+n}^3 + 54g_{12+n} q_{6+n}^2),
\label{E6discriminant}
\eeqa
showing that there is an $E_6$ singularity at $z=0$. The neutral moduli 
counting is 
\beqa
n + 7 + n + 9 + n + 13 - 1 &=& 3 n + 28,
\eeqa
which again agrees with the heterotic result. Also the heterotic prediction 
of the charge matter is $n+6$ hypermultiplets in ${\bf 27}$, which is 
certainly implied by the extra zeroes of the discriminant (\ref{E6discriminant}).

At this point, 
looking at the charged matter contents in the two examples, 
we notice an 
interesting fact: They are precisely the ones found in 
the homogeneous K\"ahler manifolds $E_8/(E_7\times U(1))$ and 
$E_7/(E_6\times U(1))$, respectively.
In fact, as shown in Table \ref{table1}, 
{\em they all correspond to a homogeneous 
K\"ahler manifold of the relevant type.}
As explained in subsequent sections, the latter are classified and labeled by 
``painted'' Dynkin diagrams \cite{BFR}. 
The corresponding diagrams are 
also shown together in Table \ref{table1}.

Why does such a relationship exist? 
A geometric explanation has been given \cite{KatzVafa} 
on how the charged matter arises at the extra singularities  
by utilizing the Cartan deformation of the singularity. 
There is, however, another alternative argument 
using string junctions \cite{Tani,CGEH_Three_looks} that is more convenient to establish the connection 
between the charged matter spectrum at an extra zero and 
a homogeneous K\"ahler manifold. 
This is the main theme of the next section.

\subsection{Matter from string junctions---Tani's argument}
We will now explain how the chiral matter spectrum is determined by 
investigating string junctions near the enhanced singularity, following \cite{Tani}
(see also \cite{CGEH_Three_looks} for a more recent discussion) .
%
%
In F-theory, singularity enhancement occurs associated with a 
singularity of an elliptic manifold on which F-theory is compactified \cite{EvidenceforFtheory}. 
Singularities of elliptic fibrations were classified according to their types 
investigated by Kodaira \cite{Kodaira}.
Technology was developed \cite{DHIZ} 
to describe these Kodaira singularities 
in terms of coalesced $[p,q]$ 7-branes and string junctions 
stretched 
between them (see \cite{GaberdielZwiebach,stringjunction1,stringjunction2,
stringjunction3,stringjunction4,stringjunction5,stringjunction6,stringjunction7,GHS1,GHS2} for more works on 
string junctions in F theory). 
We will first briefly summarize the salient features of their  
description, referring to \cite{DHIZ} for more detail.

One of the characteristic features of F-theory is that 7-branes are
allowed to change their types (=$(p,q)$ $SL(2,\ZZ)$ charges) as they wander 
about among themselves. More precisely, a 7-brane background is only 
single-valued up to $SL(2,\ZZ)$ transformations \cite{cosmic_strings}. 
Such a path-dependent 
transformation is called {\em monodromy}. For instance, as we will see in section 
\ref{SUSYdeformation},
in the single D7-brane solution the type IIB scalar $\tau$ behaves like 
$\sim \frac1{2\pi i}\log z$ near the brane locus on the 
complex $z$ plane. So if one traces the value of $\tau$ as one moves 
around the locus, $\tau$ gets transformed to $\tau+1$. 
Thus the monodromy in this case is a fractional linear transformation 
specified by 
\beqa
\left(
\begin{array}{cc}
 1 & 1     \\
 0 & 1
\end{array}
\right)&\equiv&T.
\label{T}
\eeqa

More generally, taking the convention that a $[1,0]$-brane means a D-brane,
the monodromy matrix for a $[p,q]$ brane is given by a similarity 
transformation of $T$ as
\beqa
\left(
\begin{array}{cc}
 p & r     \\
 q & s
\end{array}
\right)
\left(
\begin{array}{cc}
 1 & 1     \\
 0 & 1
\end{array}
\right)
\left(
\begin{array}{cc}
 p & r     \\
 q & s
\end{array}
\right)^{-1}
&=&\left(
\begin{array}{cc}
 1-p q & p^2 \\
 -q^2 & p q+1 \\
\end{array}
\right)\n
&\equiv&{\bf X}_{[p,q]},
\eeqa 
where $p,q,r,s$ are all integers satisfying 
$ps-qr=1$. ${\bf X}_{[p,q]}$ does not depend on 
the choice of $r$ or $s$.

What has been shown in \cite{DHIZ} is that Kodaira's 
classification of singularities of elliptic fibrations can be 
expressed by the joining/parting of several 7-branes, each of 
which is of the simplest ($I_1$) singularity type with a (relative) 
monodromy of either \footnote{In this paper we identify the labels of 
the branes (such as \AA) with its monodromy matrices. Also 
${\bf X}_{[p,q]}$ is $=K_{[p,q]}^{-1}$ in \cite{DHIZ}. Since the ordering 
of $\AA, \BB,\ldots$ is reversed for $K_A, K_B,\ldots$, this is consistent.
%
%
We should also note that the choice of \AA, {\BB}  and {\CC} branes 
does not reproduce the most general (or natural) set of vanishing cycles 
from the point of view of a deformed geometry. 
}
\beqa
\AA={\bf X}_{[1,0]}=T,~~\BB={\bf X}_{[1,-1]}=
\left(
\begin{array}{cc}
 2 & 1     \\
 -1 & 0     
\end{array}
\right)~~\mbox{or}~~
\CC={\bf X}_{[1,1]}=
\left(
\begin{array}{cc}
 0 & 1     \\
 -1 & 2     
\end{array}
\right).
\eeqa  
The correspondence is summarized in Table \ref{table2} \cite{DHIZ}.
\begin{table}
\caption{Collapsible set of 7-branes and Kodaira's 
classification \cite{DHIZ}. \label{table2}}
\centering
\begin{tabular}{|c|c|c|c|}
\hline
Fiber type&
Singularity type&
7-branes&
Brane type\\
\hline
$I_n$
&$A_{n-1}$
&$\AA^n$
&$A_{n-1}$
\\
$II$
&$A_0$
&$\AA\CC$
&$H_0$
\\
$III$
&$A_1$
&$\AA^2\CC$
&$H_1$
\\
$IV$
&$A_2$
&$\AA^3\CC$
&$H_2$
\\
$I_0^*$
&$D_4$
&$\AA^4\BB\CC$
&$D_4$
\\
$I_n^*$
&$D_{n+4}$
&$\AA^{n+4}\BB\CC$
&$D_{n+4}$
\\
$II^*$
&$E_8$
&$\AA^7\BB\CC^2$
&$E_8$
\\
$III^*$
&$E_7$
&$\AA^6\BB\CC^2$
&$E_7$
\\
$IV^*$
&$E_6$
&$\AA^5\BB\CC^2$
&$E_6$
\\
\hline 
\end{tabular}
\end{table}

\begin{figure}[h]%
\centering
\includegraphics[height=0.3\textheight]{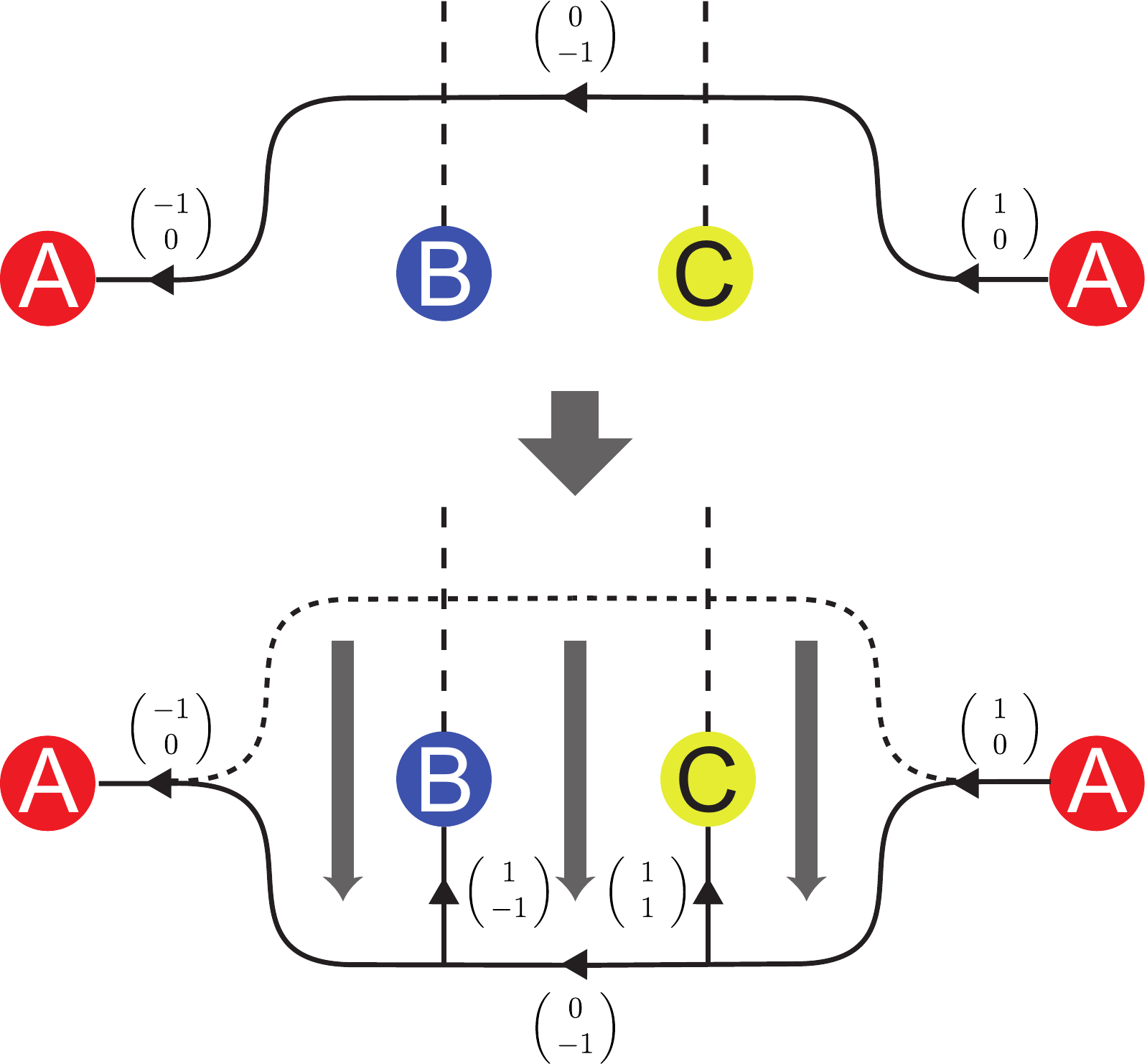}
\caption{\label{ABCA}  A string junction.}
\end{figure}

String junctions are basically the $(p,q)$ analogues of open strings. 
As before, let $[1,0]$ 7-brane be an ordinary D-brane, and let us now define
$(1,0)$ string to be the fundamental open string. Then  one can say that 
a $(1,0)$ string can end on a $[1,0]$ 7-brane. Likewise, by the $SL(2,\ZZ)$ 
S-duality, a $(p,q)$ string can end only on a $[p,q]$ 7-brane. However, as we
remarked at the beginning of this section, a $(p,q)$ string undergoes in general 
a monodromy transformation after circling around the locus of another 7-brane.
In that case the string is not of the $(p,q)$ type any more 
but becomes a different $(p',q')$ string.  
If $(p',q')$ is proportional to $(p,q)$, the string can still 
end on $[p,q]$ brane, but it creates a different state than that of the string directly 
connected between the two $[p,q]$ branes.    
For example, if a fundamental ($=(1,0)$) string circles around a \CC-brane and 
a \BB-brane, then 
\beqa
\BB \CC \left(
\begin{array}{c}
1\\ 0
\end{array}
\right)
=
\left(
\begin{array}{cc}
 2 & 1     \\
 -1 & 0     
\end{array}
\right)
\left(
\begin{array}{cc}
 0 & 1     \\
 -1 & 2     
\end{array}
\right)\left(
\begin{array}{c}1\\ 0
\end{array}
\right)=
\left(
\begin{array}{c}-1\\ 0
\end{array}
\right),
\eeqa
so it turns into a $(-1,0)$ string. This can still end on another \AA ~brane 
but its sign of the charge is inverted. A pair of \BB ~and \CC ~branes are necessary 
elements to constitute a $D$-type singularity, identified 
as an orientifold plane \cite{Ftheoryorientifold}. As shown in Figure \ref{ABCA}, there are branch cuts 
extending from the \BB  ~and \CC ~branes, and the string that experiences 
the monodromy cuts across them. But if the path of the string is deformed 
so that the 7-branes pass across the string, then the path runs in the 
region where there are no cuts, so in order for the charge conservation 
to be satisfied, some new strings with appropriate charges need to 
emerge out of the 7-branes (the Hanany-Witten effect) (Figure \ref{ABCA}). 
Such a multi-pronged 
string is called a string junction. In this example, a $(1,1)$ string is coming 
into the \CC ~brane, and a $(1,-1)$ string into the \BB ~brane. The original $(1,0)$ string 
then turns into a $(-1,0)$ string, and the charge conservation at each junction point 
is satisfied.

String  junctions are conveniently  
represented in the ``divisor-like'' form \cite{DHIZ}; if $n$ $(p,q)$ strings come  
into a $[p,q]$ 7-brane for a given multi-pronged string, then one associates 
them with 
the monomial $n {\bf x}_{[p,q]}$. Summing up these monomials over all the prongs,
one obtains the expression for the string junction as a formal sum of 
monomials with integer coefficients. For the example of Figure  \ref{ABCA}, this is
\beqa
-{\bf a}_2+{\bf b}+{\bf c}-{\bf a}_1,
\label{-a+b+c-a}
\eeqa
where ${\bf a}=x_{[1,0]}$, ${\bf b}={\bf x}_{[1,-1]}$ and ${\bf c}={\bf x}_{[1,1]}$.
On the other hand, an ordinary fundamental string directly connecting 
the two D(=\AA)-branes is represented as
\beqa
{\bf a}_2-{\bf a}_1.
\label{a-a}
\eeqa

In \cite{DHIZ}, it was proved that the 7-brane configuration
\beqa
{\bf E_N}&=&\AA^{N-1}\BB\CC^2
\eeqa
for the $E_N$ algebra for $N\geq 2$ is equivalent to
\beqa
{\bf \tilde E_N}&=&\AA^N{\bf X}_{[2,-1]}\CC
\eeqa
since they are made identical by the use of monodromy transformations 
and an $SL(2,\ZZ)$ conjugation. In fact, 
the string junctions representing the $E_8$ roots are most 
conveniently described in terms of Freudenthal's realization of $E_8$ 
\cite{Freudenthal,MizoguchiE10,MizoguchiGermar};
the exceptional Lie algebra $E_8$ is known to be generated by 
traceless 
$E^I_{~~J}$ $ (I,J=1,\ldots,9; ~~I\neq J)$ and antisymmetric tensors  
$E^{IJK}$  and 
$E^*_{IJK}$ $(1\leq I\neq J \neq K \neq I \leq 9)$ 
with the commutation relations
%
\beqa
{[}E^I_{~J},~~E^K_{~L}{]}&=&\delta^{K}_{J} E^I_{~L} -\delta^{I}_{L} E^K_{~J},\n
{[}E^I_{~J},~~E^{KLM}{]}&=&3\delta^{[M}_{I}E^{KL]I},\n
{[}E^I_{~J},~~E^*_{KLM}{]}&=&-3\delta^{I}_{[M}E^*_{KL]J},\n
{[}E^{IJK},~~E^{LMN}{]}&=&-\frac1{3!}\sum_{P,Q,R=1}^9 \epsilon^{IJKLMNPQR}E^*_{PQR},\n
{[}E^*_{IJK},~~E^*_{LMN}{]}&=&+\frac1{3!}\sum_{P,Q,R=1}^9 \epsilon_{IJKLMNPQR}E^{PQR},\n
{[}E^{IJK},~~E^*_{LMN}{]}&=&6\delta^J_{[M}\delta_N^K E^I_{~L]}~~~(\mbox{if $I\neq L,M,N$}),\n
{[}E^{IJK},~~E^*_{IJK}{]}&=&E^I_{~I}+E^J_{~J}+E^K_{~K}-\frac13\sum_{L=1}^9 E^L_{~L},
\eeqa
where $\epsilon^{123456789}=\epsilon_{123456789}=+1$.
The string junctions for the $E_8$ roots corresponding to these generators 
are  
summarized 
in Table \ref{E8junctions}.
\begin{table}
\caption{String junctions for the $E_8$ roots corresponding to generators 
in Freudenthal's realization.\label{E8junctions}}
\centering
\begin{tabular}{|cc|c|}
\hline
Generator&~~~~~&
String junction\\
\hline
$E^I_{~J}$ $(I,J=1,\ldots,8)$
&& $\aaa_I - \aaa_J$
\\
$E^{IJK}$
$(1\leq I<J<K\leq 8)$&&
$\aaa_I+\aaa_J+\aaa_K-\xxx_{[2,-1]}-\ccc$
\\
$E^*_{IJK}$
$(1\leq I<J<K\leq 8)$&&
$-(\aaa_I+\aaa_J+\aaa_K-\xxx_{[2,-1]}-\ccc)$
\\
$E^9_{~J}$ $(J=1,\ldots,8)$
&& $-\sum_{K=1}^8 \aaa_K -\aaa_J +3(\xxx_{[2,-1]}+\ccc)$
\\
$E^I_{~9}$ $(I=1,\ldots,8)$
&& $-\left(-\sum_{K=1}^8 \aaa_K -\aaa_I +3(\xxx_{[2,-1]}+\ccc)\right)$
\\
$E^{IJ9}$
$(1\leq I<J\leq 8)$&&
$-\sum_{K=1}^8 \aaa_K+\aaa_I+\aaa_J+2(\xxx_{[2,-1]}+\ccc)$
\\
$E^*_{IJ9}$
$(1\leq I<J\leq 8)$&&
$-\left(-\sum_{K=1}^8 \aaa_K+\aaa_I+\aaa_J+2(\xxx_{[2,-1]}+\ccc)\right)$
\\
\hline
\end{tabular}
\end{table}

The fact that the Kodaira singularities are described by coinciding 
7-branes, and the existence of varieties of string junctions which 
correspond to the roots of the exceptional group, offer a natural 
explanation \cite{GaberdielZwiebach} 
for the origin of the exceptional group gauge symmetry 
in F-theory. 
When $N$ D-branes come on top of each other, one 
gets $U(N)$ gauge symmetry \cite{Witten_bound_states}. In this 
case the relevant massless ``W-bosons'' are supplemented by 
the excitations of light open strings ending on different D-branes. 
Likewise, the extra massless fields needed for the gauge symmetry 
enhancement to an exceptional group can be thought of as 
coming from the string junctions connecting the collapsing 7-branes 
nontrivially \cite{GaberdielZwiebach}.

\begin{figure}[h]%
\centering
\centerline{\includegraphics[height=0.4\textheight]{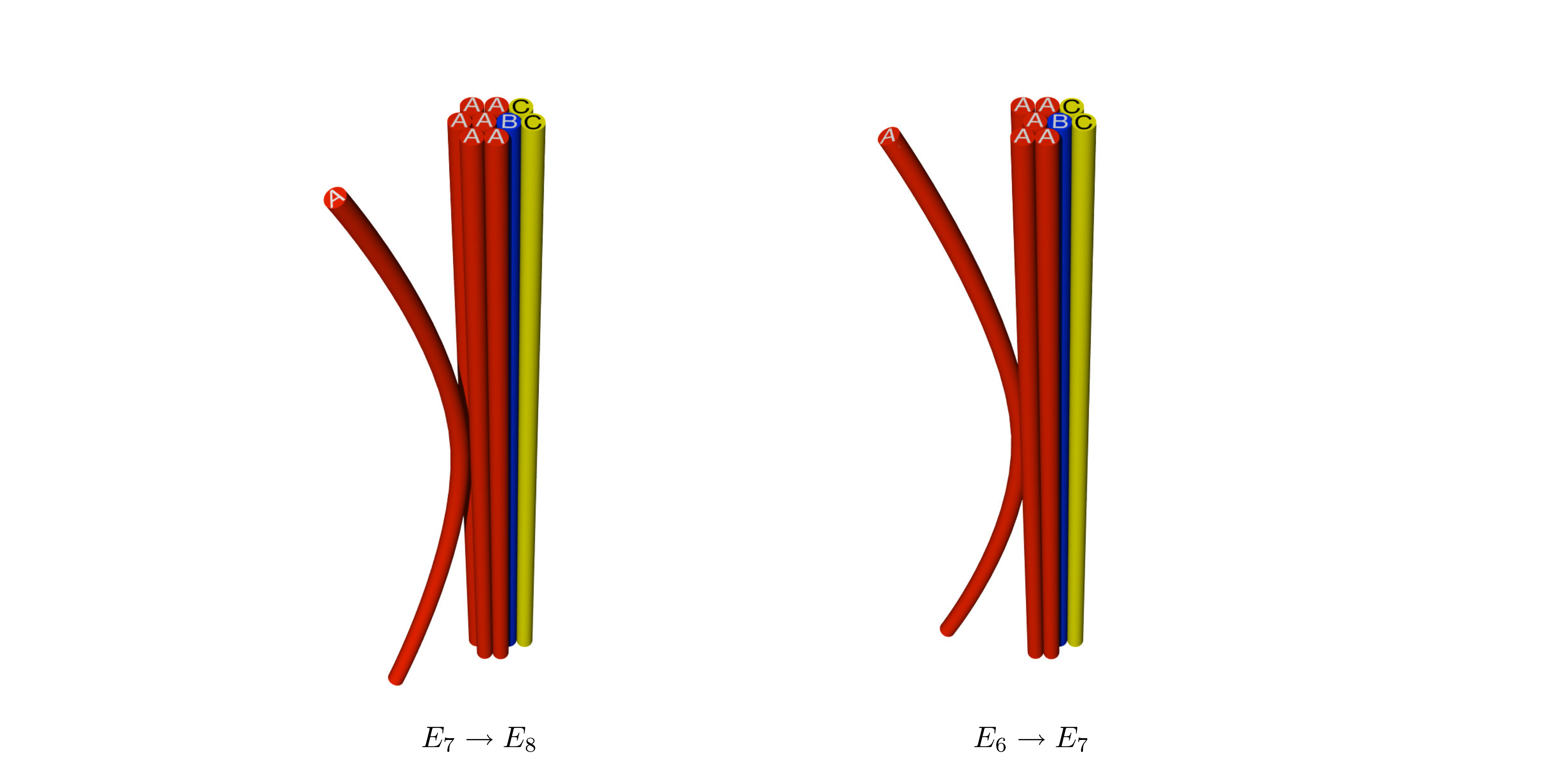}}
\caption{\label{E8toE7E7toE6}  Extra zeroes and bending branes.
Left: The $E_7$ singularity is enhanced to $E_8$ at the extra zero. 
Right: Similarly the $E_6$ is enhanced to $E_7$.}
\end{figure}

Having reviewed the 7-brane technology, we are now in a position 
%
%
to explain the argument by Tani on the emergence of 
matter 
at the extra zeroes in terms of string junctions. As we saw before, 
the singularity of elliptic fibers is more enhanced at an extra zero 
than elsewhere around the point. In the 7-brane picture, this means 
that another 7-brane comes into join the bunch of coincident 
7-branes to meet at that point. This is illustrated in Figure \ref{E8toE7E7toE6}.

The string junctions at the extra zero are divided into two 
different classes. The junctions which do not have an end on the 
bending brane can move apart from the extra zero without any loss 
of energy, and hence create a massless gauge multiplet similarly to the above. 
Let {\myeu h} be the Lie algebra of this gauge multiplet. 
On the other hand, those which {\em do} have an end on the bending brane 
cannot move away from there 
but localized near that point. 
Let {\myeu g} be the Lie algebra of the enhanced singularity 
at the extra zero, then they correspond to the elements of {\myeu g} 
that do not belong to  {\myeu h}$\oplus U(1)$.
They consist of a pair of representations of {\myeu h} that are 
complex conjugate to each other, so the states they create must be hypermultiplets.
But since a half of the supersymmetries are broken,
these states do not necessarily have
the same mass
but it is possible that only a half of them remains massless. 
(The condition for half of the SUSY to be preserved will be 
discussed in detail in section \ref{SUSYdeformation}.)
In the intersecting D-brane systems, where the explicit quantization is 
possible, 
this is a familiar phenomenon \cite{BDL}.
So if one assumes that this is also true in the general case when
 the singularity contains not only D-branes but other $[p,q]$ branes, 
 then one can get 
a perfectly consistent picture of the chiral matter generation \cite{Tani}. 
The relevant set of coalesced 7-branes and and the extra brane(s) 
coming to join are shown in Table \ref{table1} 
for each case of unbroken gauge symmetry. 


\begin{figure}[b]%
\centering
\centerline{
\includegraphics[height=0.4\textheight]{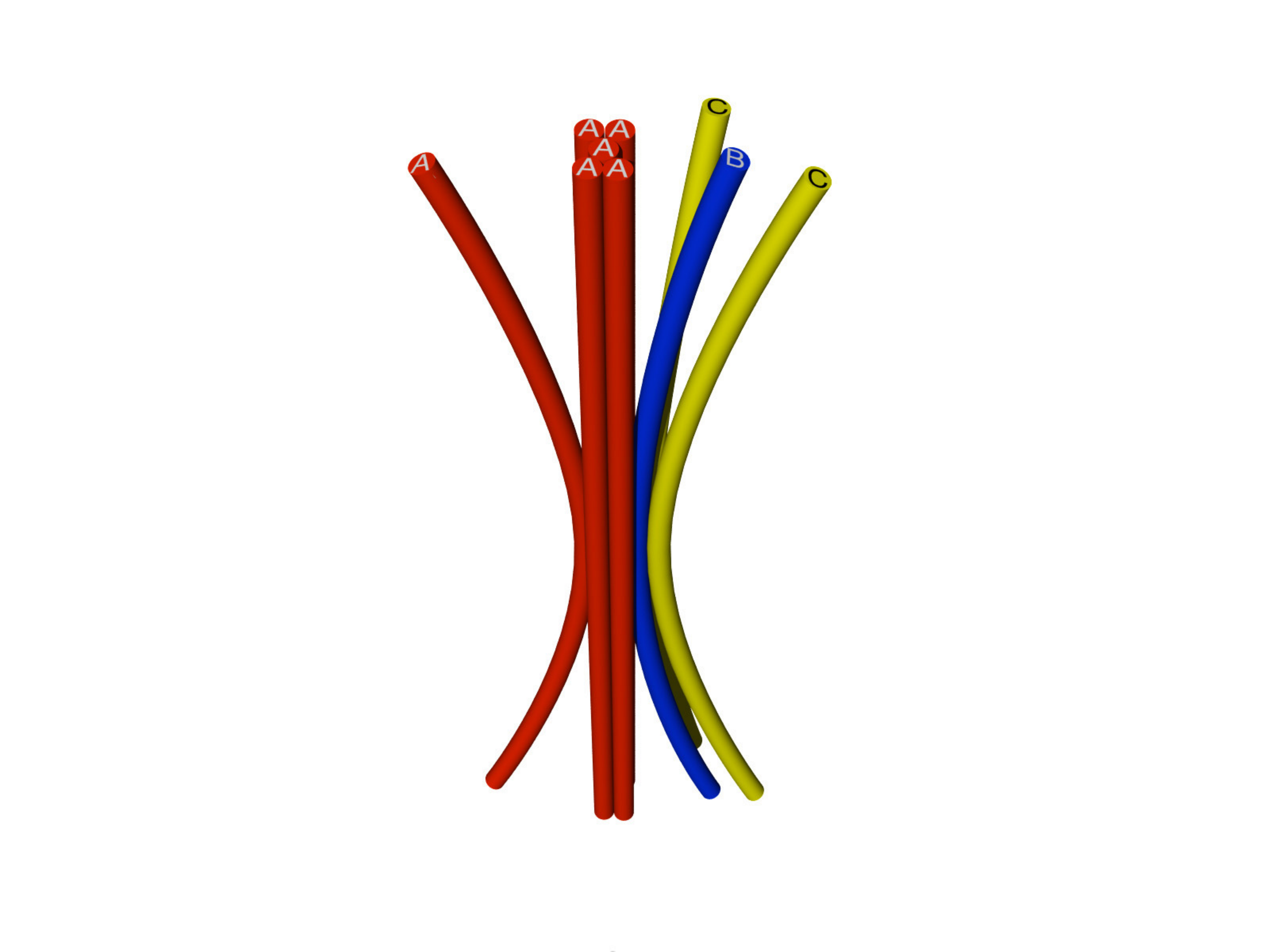}
}
\caption{\label{spaghetti}  The 7-brane configuration for the $E_7/(SU(5)\times U(1)^3)$ model.}
\end{figure}

\subsection{The correspondence to homogeneous K\"ahler manifolds 
and 
the $E_7/(SU(5)\times U(1)^3)$ model
}\label{subsection:The_correspondence}
In particular,
it can readily explain why the coset structure arises at the extra zeroes.
For example, on the left of Figure\ref{E8toE7E7toE6}, 
the string junctions localized at the 
intersection span $E_8$, and among them only those corresponding 
to $E_7$ can freely move along the bunch of coalesced branes.  
Thus the states created by the junctions with an end on the bending \AA~brane are
in $E_8$ but not in the $E_7$ subalgebra of it. Taking either of the complex conjugate pair,
one gets  $E_8/(E_7\times U(1))$ 
and hence a {\bf 56} of $E_7$.
Similarly, on the right of Figure \ref{E8toE7E7toE6}, the junctions at the intersection point give $E_7$, 
and those do not have a leg on the \CC~brane are the $E_6$ part of it. 
Taking a half of the rest, one obtains 
$E_7/(E_6\times U(1))$, that is,  a {\bf 27} of $E_6$.

At this point it is now obvious what 7-brane configuration would yield 
the spectrum of the Kugo-Yanagida $E_7/(SU(5)\times U(1)^3)$ model.
%
All we need is a set of 
six \AA~branes, 
one \BB~brane and two \CC~branes, such that at a generic position 
along some complex dimension ($w$) only five of the \AA~branes are collapsed   
while all the other branes are apart, but at a certain juncture point they all join 
together (Figure \ref{spaghetti}).
If such a brane configuration exists and preserves half of the supersymmetries, 
then the string junctions that have an end on the bending \AA~brane will produce 
six-dimensional hypermultiplets transforming in 
${\bf 10}\oplus {\bf 5}\oplus {\bf \bar 5}\oplus {\bf \bar 5}\oplus {\bf 1}\oplus {\bf 1}$
of $SU(5)$ from {\bf 27} of $E_6$,  the junctions with an end on one of the \CC~branes  
(but not on the \AA~brane bending away from the juncture) 
will create  ${\bf 10}\oplus {\bf \bar 5}\oplus {\bf 1}$ 
from  {\bf 16} of $SO(10)$, and those ending on the remaining \BB~ and \CC~ branes 
(not connected to the \AA~ and \CC~ branes above) will yield ${\bf 10}$ 
of $SU(5)$. They have exactly the same quantum numbers as those 
of the $E_7/(SU(5)\times U(1)^3)$ model, except that they are still 
six-dimensional supermultiplets.  So in order to get the 
four-dimensional ${\cal N}=1$ family unification model 
we will need to further compactify two of the  
spatial dimensions, reduce the number of 
supersymmetries and project out one half of the four-dimensional chiral multiplets 
with either of the chiralities. But before we discuss how to do that 
we would like to consider the more fundamental question: Can the 7-brane configuration 
as described above 
preserve SUSY? 
This is because, if not, our proposal of how to achieve supersymmetric 
family unification in F-theory would break down.  
Fortunately, however, we can prove that it indeed can, as we show 
in the next section.



\section{Holomorphic deformation of 7-branes}
\label{SUSYdeformation}
\subsection{The stringy-cosmic-string solution
}
Let us first recall the 7-brane solution of type IIB supergravity 
on $\PP^1$ obtained long time ago by Greene et.al.\cite{cosmic_strings}.
Although this is also a well-known material, we will take a different route to it, 
examining   
the SUSY conditions;
our approach is 
more convenient and can be directly extended to the deformed case discussed 
in the next section.

The metric ansatz is
\beqa
ds^2&=&-dt^2 + e^{\varphi(z,\bar z)}dz d\bar z + (dx^i)^2~~~(i=1,\ldots,7).
\eeqa
The type IIB complex scalar field 
$
\tau=C_0+ie^{-\phi},
$
where $C_0$ and $\phi$ are the RR scalar and the dilaton, respectively,
is assumed to be a holomorphic function depending only on $z$:
\beqa
\tau&=&\tau(z).
\eeqa 
The other supergravity fields are set to zero. In this case the supersymmetry 
variations of the gravitino and the dilatino are \cite{HoweWest,Schwarz}
\beqa
\delta\psi_\mu&=&\frac 1\kappa\left(
\partial_\mu -
\frac14 \omega_{\mu\alpha\beta}\gamma^{\alpha\beta}-\frac i2Q_\mu
\right)\epsilon,\label{gravitino_variation}
\\
\delta\lambda&=&\frac i\kappa P_\mu \gamma^\mu \epsilon^*,
\label{dilatino_variation}
\eeqa
where $P_\mu$ and $Q_\mu$ are well-known $SU(1,1)$-invariant 
connections given by
\beqa
P_\mu&=&-\frac{\partial_\mu \tau}{\tau-\bar\tau},\\
Q_\mu&=&-\frac i2 \frac{\partial_\mu(\tau + \bar\tau)}{\tau - \bar\tau}.
\eeqa

Since we have assumed that $\tau$ is holomorphic, we have
\beqa
P_{\bar z}=0.
\eeqa
Let $z$, $\bar z$ be complex linear combinations of real coordinate 
$x^{\dot 8}$ and $x^{\dot 9}$ :\footnote{The dots indicate that those indices 
are curved ones.}
\beqa
z=x^{\dot 8}+i x^{\dot 9},~~~\bar z=x^{\dot 8}-i x^{\dot 9}
\eeqa
and take the two-dimensional gamma matrices
\beqa
\gamma^8=\sigma_1,~~~  \gamma^9=\sigma_2,  
\eeqa 
where each component is understood to act on
spinors in the eight-dimensional space-time. Then 
\beqa
\delta\lambda&\propto&P_z\gamma^z\epsilon^*\n
&=&\left(
\begin{array}{cc}0&*
\\0&0
\end{array}
\right)\epsilon^*,
\eeqa
so $\delta\lambda$ vanishes for $\epsilon$ of the form 
$\left(
\begin{array}{c}*
\\0
\end{array}
\right)$.

On the other hand, the gravitino variation $\delta\psi_\mu$ depends 
on the spin connections and $Q_\mu$.
The only nontrivial components of the spin connection are
\beqa
\omega_{z89}&=&\frac i2 \partial_z \varphi~=~-\omega_{z98},\\
\omega_{\bar z89}&=&-\frac i2 \partial_{\bar z} \varphi~=~-\omega_{\bar z98},
\eeqa
while 
\beqa
Q_z&=&-\frac i2 \frac{\partial_z\tau}{\tau - \bar\tau}~=~
-\frac i2 \frac{\partial_z(\tau - \bar\tau)}{\tau - \bar\tau},\n
Q_{\bar z}&=&-\frac i2 \frac{\partial_{\bar z}{\bar \tau}}{\tau - \bar\tau}~=~
+\frac i2 \frac{\partial_{\bar z}{(\tau - \bar\tau)}}{\tau - \bar\tau}
\eeqa
since $\tau$ is assumed to be holomorphic. Let
\beqa
\epsilon&=&\left(
\begin{array}{c}
\tilde\epsilon
\\0
\end{array}
\right),
\eeqa
then
\beqa
\left(
\partial_z -\frac14 \omega_{z\alpha\beta}\gamma^{\alpha\beta}
-\frac i2 Q_z
\right)\epsilon
&=&
\left(
\begin{array}{c}
\partial_z\tilde\epsilon
+\frac14
\partial_z(\varphi-\log(\tau-\bar\tau))
\cdot
\tilde\epsilon
\\0
\end{array}
\right),
\n
\left(
\partial_{\bar z} -\frac14 \omega_{\bar z\alpha\beta}\gamma^{\alpha\beta}
-\frac i2 Q_{\bar z}
\right)\epsilon
&=&
\left(
\begin{array}{c}
\partial_{\bar z}\tilde\epsilon
-\frac14
\partial_{\bar z}(\varphi-\log(\tau-\bar\tau))
\cdot
\tilde\epsilon
\\0
\end{array}
\right).
\eeqa
Therefore, if $\varphi=\log\frac{\tau-\bar\tau}{2i}$, a Killing spinor (which is 
constant in this case)  
exists 
and half the supersymmetries are preserved. More generally, if
\beqa
\varphi&=&\log\frac{\tau-\bar\tau}{2i}+F(z)+\bar F(\bar z)
\eeqa
for some holomorphic function $F(z)$, then
$\epsilon$ with 
\beqa
\tilde\epsilon&=&e^{\frac14(F-\bar F)}\times \mbox{const.}
\eeqa
is a Killing spinor. $F(z)$ is chosen to be \cite{cosmic_strings}
\beqa
F(z)&=&2\log\eta(\tau(z))+f(z). \label{F(z)}
\eeqa
The first term is for the modular (or S-duality) invariance, 
while $f(z)$ is some function of $z$ to compensate the zeroes at the 
7-brane loci. For instance, for a single D7-brane at $z=0$ we have locally 
\cite{cosmic_strings}
\beqa
\tau\sim\frac1{2\pi i}\log z,~~~f(z)\sim-\frac1{12}\log z.
\eeqa

\subsection{A supersymmetric deformation of 7-branes}
In the previous section we have reviewed the 7-brane solutions in eight 
dimensions---in modern terminology this is a codimension-one singularity.
We now turn to a codimension-two singularity, that is, 
we deform $\tau$ so that it also varies over another holomorphic coordinate 
$w=x^{\dot 6}+ i x^{\dot 7}$:
\beqa
\tau&=&\tau(z,w).
\label{holomorphic_tau}
\eeqa
Note 
that we do not specify any particular 
 {\em global} geometry.
We will show that, for any 
such holomorphic deformation 
\footnote{up to some 7-brane loci where $\tau$ diverges 
logarithmically.}
of the modulus function $\tau$, there exists, 
at least {\em locally}, some K\"{a}hler metric such that it preserves a quarter 
of supersymmetries.

We focus on the four-dimensional part of the ten-dimensional metric,
which we assume to be hermitian:
\beqa
ds_4^2&=&e^{\Phi}dzd\bar z +e^{\Psi}(dw + \xi dz)(d \bar w + \bar \xi d\bar z).
\eeqa
Any hermitian metric can be written in this form with two real functions 
$\Phi$, $\Psi$ and a complex function $\xi$. The vierbein of this subspace 
is block diagonal:
\beqa
e_\mu^{~\alpha}&=&\left(
\begin{array}{cc}
e_i^{~a}&0\\
0&e_{\bar i}^{~{\bar a}}
\end{array}
\right),
\eeqa
where $\mu=i,\bar i$; $i=z,w$; $\bar i=\bar z,\bar w$; 
$\alpha=a,\bar a$; $a=1,2$; $\bar a= \bar 1, \bar 2$ with
\beqa
e_i^{~a}&\equiv&\left(
\begin{array}{cc}
e_i^{~8}+ie_i^{~9}&e_i^{~6}+ie_i^{~7}\\
\end{array}
\right)
~=~\left(
\begin{array}{cc}
e^{\frac\Phi 2}&e^{\frac\Psi 2} \xi \\
0&e^{\frac\Psi 2}
\end{array}
\right),\n
e_{\bar i}^{~\bar a}&\equiv&\left(
\begin{array}{cc}
e_{\bar i}^{~8}-ie_{\bar i}^{~9}&e_{\bar i}^{~6}-ie_{\bar i}^{~7}\\
\end{array}
\right)
~=~\left(
\begin{array}{cc}
e^{\frac\Phi 2}&e^{\frac\Psi 2} \bar\xi \\
0&e^{\frac\Psi 2}
\end{array}
\right).
\eeqa
In our convention the flat metric is
\beqa
\eta_{\alpha\beta}&=&\left(
\begin{array}{cc}
&\frac12~\! \II\\
\frac12~\! \II&
\end{array}
\right),~~~\II=
\left(
\begin{array}{cc}
  1~&  0  \\
  0~&  1
\end{array}
\right),
\eeqa
so that
\beqa
g_{\mu\nu}&=&e_\mu^{~\alpha}\eta_{\alpha\beta}e_{~\nu}^{\beta},~~~
ds_4=g_{\mu\nu}dx^\mu dx^\nu.
\eeqa

We take 
\beqa
&&\gamma^8=\sigma_1\otimes \II,~~~\gamma^9=\sigma_2\otimes \II \n
&&\gamma^6=\sigma_3\otimes \sigma_1,~~~\gamma^7=\sigma_3\otimes \sigma_2,
\eeqa
so that
\beqa
\gamma^1&\equiv&\gamma^8+i \gamma^9
~=~
\left(
\begin{array}{cc}
  &   ~ ~2 \\
0  &     
\end{array}
\right)
\otimes \II
~=~
\mbox{\scriptsize $\left(
\begin{array}{cccc}
  &   &  2& \\
  &   &   &2\\
0  &   &   &\\
  &  0 &   &
\end{array}
\right)$},
\n
\gamma^{\bar 1}&\equiv&\gamma^8-i \gamma^9
~=~
\left(
\begin{array}{cc}
  &   ~ ~0 \\
2  &     
\end{array}
\right)
\otimes \II
~=~
\mbox{\scriptsize $\left(
\begin{array}{cccc}
  &   &  0& \\
  &   &   &0\\
2  &   &   &\\
  &  2 &   &
\end{array}
\right)$},
\n
\gamma^2&\equiv&\gamma^6+i \gamma^7
~=~
\sigma_3
\otimes \left(
\begin{array}{cc}
  &   ~ ~2 \\
0  &     
\end{array}
\right)
~=~
\mbox{\scriptsize $\left(
\begin{array}{cccc}
  &   ~2&  & \\
  0&   &   &\\
  &   &   &-2\\
  &   &   0&
\end{array}
\right)$},
\n
\gamma^{\bar 2}&\equiv&\gamma^6-i \gamma^7
~=~
\sigma_3
\otimes \left(
\begin{array}{cc}
  &   ~ ~0 \\
2  &     
\end{array}
\right)
~=~
\mbox{\scriptsize $\left(
\begin{array}{cccc}
  &   ~0&  & \\
  2&   &   &\\
  &   &   &0\\
  &   &   -2&
\end{array}
\right)$}.
\label{gamma_matrices}
\eeqa

Due to the holomorphic assumption (\ref{holomorphic_tau}), we have, again,
\beqa
P_{\bar i}=0~~~(\bar i=\bar z, \bar w). 
\eeqa
The 
dilatino variation thus reads 
\beqa
\delta\lambda&\propto&P_i e_a^{~i} \gamma^a \epsilon^*.
\eeqa
Since the leftmost columns of 
$\gamma^a$ $(a=1,2)$ are zero as displayed in (\ref{gamma_matrices}), 
 $\delta\lambda$ vanishes for a SUSY variation parameter
of the form
\beqa
\epsilon&=&
\mbox{\scriptsize $\left(
\begin{array}{c}
\tilde\epsilon\\
0\\
0\\
0
\end{array}
\right)$}.
\label{SUSY_parameter}
\eeqa

We will now examine under what conditions the gravitino variation 
$\delta \psi_\mu$ also vanishes for 
$\epsilon$ (\ref{SUSY_parameter}). Since the nonzero component 
is only the first one, we are only concerned with the first columns of 
$\omega_{\gamma\alpha\beta}\gamma^{\alpha\beta}$: 
\beqa
\omega_{1\alpha\beta}\gamma^{\alpha\beta}&=&
\left(
\begin{array}{cccc}
-e^{-\frac\Phi 2}(\partial_w\xi -\xi\partial _w\Phi + \partial_z \Phi)&*&*&*\\
0&*&*&*\\
0&*&*&*\\
2e^{-\Phi -\frac\Psi 2}\left(e^\Psi(\bar\xi\partial_{\bar w}\xi-\partial_{\bar z}\xi)
+e^\Phi \partial_{\bar w}\Phi\right)
&*&*&*
\end{array}
\right),\n
\omega_{2\alpha\beta}\gamma^{\alpha\beta}&=&
\left(
\begin{array}{cccc}
e^{-\frac\Psi 2}\left( e^{\Psi - \Phi}
(\xi\partial_{ w}\bar \xi  - \partial_z \bar\xi) - \partial_w \Psi
\right)&*&*&*\\
0&*&*&*\\
0&*&*&*\\
2e^{-\frac\Phi 2}(\partial_{\bar w}\bar\xi +\bar\xi\partial_{\bar w} \Psi - \partial_{\bar z} \Psi)
&*&*&*
\end{array}
\right),\n
\omega_{\bar 1\alpha\beta}\gamma^{\alpha\beta}&=&
\left(
\begin{array}{cccc}
-(\overline{\mbox{$(1,1)$ component of $\omega_{1\alpha\beta}\gamma^{\alpha\beta}$}})&*&*&*\\
0&*&*&*\\
0&*&*&*\\
0&*&*&*
\end{array}
\right),
\n
\omega_{\bar 2\alpha\beta}\gamma^{\alpha\beta}&=&
\left(
\begin{array}{cccc}
-(\overline{\mbox{$(1,1)$ component of $\omega_{2\alpha\beta}\gamma^{\alpha\beta}$}})&*&*&*\\
0&*&*&*\\
0&*&*&*\\
0&*&*&*
\end{array}
\right).
\label{omega_gamma}
\eeqa
Since the ``Bismut-like'' connection (\ref{gravitino_variation}) contains,
besides the spin (=Levi-Civita) connection, only $Q_\mu$ which is $U(1)$, the 
gravitino variations vanish only if the off-diagonal components (of 
the first columns) do. Taking their complex conjugates, the conditions 
are
\beqa
e^\Psi(\xi\partial_{w}\bar\xi-\partial_{z}\bar\xi)
+e^\Phi \partial_{w}\Phi&=&0~~~\mbox{and}
\label{cond1}\\
\partial_{w}\xi +\xi\partial_{w} \Psi - \partial_{z} \Psi
&=&0.
\label{cond2}
\eeqa
It is easy to see that they are equivalent to
\beqa
\partial_w(e^\Psi\xi\bar\xi + e^\Phi)&=&
\partial_z(e^\Psi\bar\xi)~~~ \mbox{and}\\
\partial_w(e^\Psi\xi)&=&\partial_z e^\Psi,
\eeqa
or
\beqa
\partial_i g_{j\bar i}=\partial_j g_{i\bar i},~~~
\partial_{\bar i} g_{\bar j i}=\partial_{\bar j} g_{\bar i i},
\eeqa
which are satisfied if the metric is K\"{a}hler. Conversely, if the conditions 
(\ref{cond1}),(\ref{cond2}) are satisfied, then the spin connection
turns out to be a $U(2)$ connection and hence the metric is K\"{a}hler.

Suppose that we have a solution to the conditions (\ref{cond1}),(\ref{cond2}). 
Such a solution exists, at least locally, as we will show in a moment. 
Then using 
them in (\ref{omega_gamma}), we find
\beqa
\omega_{i\alpha\beta}\gamma^{\alpha\beta}=
\left(
\begin{array}{cccc}
-\partial_i(\Phi + \Psi)
&*&*&*\\
0&*&*&*\\
0&*&*&*\\
0&*&*&*
\end{array}
\right),
~~~
\omega_{\bar i\alpha\beta}\gamma^{\alpha\beta}=
\left(
\begin{array}{cccc}
+\partial_{\bar i}(\Phi + \Psi)
&*&*&*\\
0&*&*&*\\
0&*&*&*\\
0&*&*&*
\end{array}
\right).
\eeqa
On the other hand, $Q_\mu$'s are given by
\beqa
Q_i&=&-\frac i2 \partial_i\log(\tau - \bar\tau),\n
Q_{\bar i}&=&+\frac i2 \partial_{\bar i}\log(\tau - \bar\tau).
\eeqa
Therefore, similarly to the previous section, we have a Killing spinor if
\beqa
\Phi + \Psi &=& \log\frac{\tau - \bar\tau}{2i} + F(z^i) + \bar F({\bar z}^{\bar i})
\label{Phi+Psi}
\eeqa
for some holomorphic function $F$ of $z^i= z,w$.  Again, we can set 
$F(z)$ similarly to (\ref{F(z)}) for some $f(z)$ which compensates 
the zeroes of the brane loci, then the metric is positive and modular 
invariant.

Now what remains to be done is to show the existence of 
$\Phi$, $\Psi$ and $\xi$ that satisfy (\ref{cond1}), (\ref{cond2})
with the constraint (\ref{Phi+Psi}) 
for a given $\tau(z^i)$.
First we note that, if $\tau$ is only a function of $z$ and does not 
depend on $w$, then the problem reduces to that discussed 
in the previous section with
\beqa
\Phi=\varphi,~~~\Psi=\xi=0,
\eeqa
which obviously satisfy (\ref{cond1}), (\ref{cond2}) and (\ref{Phi+Psi}). 
So for given 
\beqa
\tau(z,w)&=&\sum_{n,\bar n =0}^\infty \tau_{(n,\bar n)}(z)w^n {\bar w}^{\bar n}\n
F(z,w)&=&\sum_{n,\bar n =0}^\infty F_{(n,\bar n)}(z)w^n {\bar w}^{\bar n},
\eeqa
we determine 
\beqa
\Phi(z,w)&=&\sum_{n,\bar n =0}^\infty \Phi_{(n,\bar n)}(z)w^n {\bar w}^{\bar n}\n
&=&\Phi_{(0,0)}(z)+\Phi_{(1,0)}(z) w+ \Phi_{(0,1)}(z) \bar w
+\Phi_{(1,1)}(z) w\bar w +\cdots,\n
\Psi(z,w)&=&\sum_{n,\bar n =0}^\infty \Psi_{(n,\bar n)}(z)w^n {\bar w}^{\bar n}\n
&=&\Psi_{(0,0)}(z)+\Psi_{(1,0)}(z) w+ \Psi_{(0,1)}(z) \bar w
+\Psi_{(1,1)}(z) w\bar w +\cdots,\n
\xi(z,w)&=&\sum_{n,\bar n =0}^\infty \xi_{(n,\bar n)}(z)w^n {\bar w}^{\bar n}\n
&=&\xi_{(0,0)}(z)+\xi_{(1,0)}(z) w+ \xi_{(0,1)}(z) \bar w
+\xi_{(1,1)}(z) w\bar w +\cdots,
\eeqa
with
\beqa
\Psi_{(0,0)}(z)=\xi_{(0,0)}(z)=0.
\eeqa
To do this 
it is more convenient to 
%
write
\beqa
e^\Psi=A, ~e^\Psi \xi=B, ~e^\Psi \bar\xi = \bar B.
\eeqa
$A$ is a real, while $B$ is a complex function 
of $z,w,\bar z$ and $\bar w$. Then
\beqa
e^\Psi \xi\bar\xi + e^\Phi&=&e^{-\Psi}(B\bar B + e^{\Phi + \Psi})
\n
&=&
A^{-1}(B\bar B +\frac{\tau - \bar\tau}{2i}e^{F+ \bar F}), 
\eeqa
where we have used (\ref{Phi+Psi}) in the last line.
Let us also set
\beqa
\frac{\tau - \bar\tau}{2i}e^{F+ \bar F}&=&h(z,w,\bar z,\bar w),
\eeqa
where the real function $h(z,w,\bar z,\bar w)$ is determined 
by the given holomorphic functions $\tau$ and $F$.  
The system of equations is now
\beqa
\partial_w B&=& \partial_z A,
\label{system_of_eq1}
\\
\partial_w(A^{-1}(B\bar B + h))&=&\partial_z\bar B,
\label{system_of_eq2}
\eeqa
which can be solved by iteration.
We expand
\beqa
A&=&\sum_{n,\bar n=0}^\infty w^n \bar w^{\bar n} A_{(n,\bar n)}(z,\bar z),\\
h&=&\sum_{n,\bar n=0}^\infty w^n \bar w^{\bar n} h_{(n,\bar n)}(z,\bar z),\\
B&=&\sum_{
\mbox{\scriptsize $
\begin{array}{c}
    n,\bar n=0  \\
    (n,\bar n)\neq (0,0)
\end{array}
$}
}^\infty w^n \bar w^{\bar n} B_{(n,\bar n)}(z,\bar z),\\
\bar
B&=&\sum_{
\mbox{\scriptsize $
\begin{array}{c}
    n,\bar n=0  \\
    (n,\bar n)\neq (0,0)
\end{array}
$}
}^\infty w^n \bar w^{\bar n} \bar B_{(n,\bar n)}(z,\bar z).
\eeqa
Since $A$ is real, we have
\beqa
\overline{A_{(n,\bar n)}}&=&A_{(\bar n,n)},
\label{AbarA}
\eeqa
whereas
\beqa
\overline{B_{(n,\bar n)}}&=&{\bar B}_{(\bar n,n)}.
\label{BbarB}
\eeqa
Plugging the expansions in (\ref{system_of_eq1}) and
(\ref{system_of_eq2}), we find
\beqa
nB_{(n,\bar n)}&=&\partial_z A_{(n-1,\bar n)},
\label{nBdelA}\\
n\left(
A^{-1}(B\bar B + h)
\right)_{(n,\bar n)}&=&\partial_z\bar B_{(n-1,\bar n)},
\label{nAinvBBh}
\eeqa
where, as obviously, $\left(
A^{-1}(B\bar B + h)
\right)_{(n,\bar n)}$ is the coefficient of $w^n \bar w^{\bar n}$ 
in the expansion of $A^{-1}(B\bar B + h)$.
Using (\ref{BbarB}),(\ref{nBdelA}) and (\ref{AbarA}), (\ref{nAinvBBh})
is further written as
\beqa
n\bar n\left(
A^{-1}(B\bar B + h)
\right)_{(n,\bar n)}&=&
\partial_z \partial_{\bar z}A_{(n-1, \bar n-1)}.
\label{nAinvBBh2}
\eeqa
Using (\ref{nBdelA}) and (\ref{nAinvBBh}) with the initial conditions
\beqa
A_{(0,0)}=1,~~B_{(0,0)}=\bar B_{(0,0)}=0
\eeqa 
and arbitrary 
functions $B_{(0,\bar n)}$ $(\bar n=1,2,\ldots)$,
$A_{(n, \bar n)}$ and $B_{(n,\bar n)}$ (and hence $\bar B_{(n, \bar n)}$) can be
determined iteratively. 
Thus we have shown that the equations (\ref{cond1}) and (\ref{cond2})
with (\ref{Phi+Psi}) have a solution, at least locally near $z=w=0$. This 
completes the proof of the existence of a local supersymmetric solution 
for any given holomorphic complex scalar function $\tau(z,w)$. 

The explicit forms of $A_{(n, \bar n)}$ obtained as a result of the
iteration are given in Appendix A up to $n+\bar n\leq 3$. 

Finally, since the $J$ function is also holomorphic, holomorphic 
deformations of the coefficient functions in the Weierstrass form 
lead to a local supersymmetric solution of type IIB supergravity.

\section{Orbifolds and anomalies}
\label{Orbifolds}
%

Thus the 7-branes described in \ref{subsection:The_correspondence} 
(the ones like a bunch of raw spaghetti) preserve SUSY. So if we 
compactify two more dimensions to reduce the SUSY to ${\cal N}=1$ 
and drop one half of the chiral supermultiplets with a definite chirality,
then we end up with precisely the $E_7/(SU(5)\times U(1)^3)$ 
supersymmetric family unification model. 
This could be done 
in a variety of ways.  The most modern way to do this is to turn on (after 
compactifying to four dimensions) 
%
%
``G-fluxes'' or
some appropriate vortex 
Higgs field \cite{DonagiWijnholt}. Then depending on the sign 
of the charge, similarly to the $Y$-charge explained 
in section \ref{CosetSpaceFamilyUnification}, a half 
of the solutions of the Dirac equation would become nonnormalizable and 
one is left with a chiral spectrum.
Alternatively, one could simply compactify two of the six dimensions 
on a two torus $T^2$, wrap the 7-branes around it and take an  
orbifold (see e.g.\cite{orbifoldGUT,6DorbifoldGUT} and references therein) 
to reduce the SUSY, thereby imposing a boundary condition 
such that the massless fields with, say, positive $Y$-charges are  
projected out. This would lead to the same effect as turning on a Higgs 
field, and hence is an effective and efficient way.  
It will be a bit ad hoc, but we emphasize that 
the necessary set of fields with correct gauge quantum numbers 
are already fixed before the projection; there is no 
need to adjust anything but is only need to eliminate a half to get them. 
More concrete discussion on this part of the construction will be 
given elsewhere.

There is one more thing to be discussed at this point: The $E_7/(SU(5)\times H)$  
models are anomalous, in the senses of both the gauge anomaly 
and the sigma-model anomaly. This problem has been for a long 
time \cite{YanagidaYasui}:  The $E_7/(SU(5)\times H)$ model includes 
a single {\bf 5} representation, and there is nothing else in the sigma model 
itself to cancel its anomaly. 

In six dimensions there is no problem; 
the six-dimensional heterotic spectrum on K3 is of course 
anomaly-free, and so is it for F-theory. If one focuses on a particular 
extra zero point, then the anomaly balance will be lost, but upon compactification 
to four dimensions, they become non-chiral and hence have no anomaly.   
Therefore, this is the issue only after the chiral projection. 

There are at least two ways out of this problem. One is the idea that 
there arises some extra matter from the orbifold. This idea has already 
been pointed out by several authors. Another, more interesting possibility 
is that extra anomalous contribution to the effective action on the brane 
might come in from the bulk, known as the anomaly inflow 
mechanism \cite{anomalyinflow1,anomalyinflow2,anomalyinflow3},
\cite{anomalyinflow5,KM,anomalyinflow7,CGEH_Three_looks}. This is an interesting possibility, but 
if this is true, then it would lead to a non-trivial prediction in the Higgs 
sector, since it is not  
the quantum effect of some ${\bf \bar 5}$ field, but some other 
effective contribution induced from the bulk, 
that cancels the anomaly of the {\bf 5} field of the model.
We postpone a more concrete analysis to future work.

\section{The explicit expression for the curve of the brane configuration
}
\label{ExplicitCurve}
Going back to the brane configuration in 
section \ref{subsection:The_correspondence},
we present in this section the 
explicit local expression for the curve that represents the 
brane configuration as shown in Figure \ref{spaghetti} 
discussed at the end of section \ref{section3},
which is to realize (after the compactification and projection)  
the $E_7/(SU(5)\times U(1)^3)$ model.
As we have shown in the last section, there exists a local 
supersymmetric solution for a holomorphically varying 
scalar function $\tau(z,w)$. Therefore, since the $J$ function is holomorphic, 
we have only to consider the Weierstrass form
\beqa
y^2&=&x^3+f(z,w) x + g(z,w)
\eeqa
for holomorphic functions $f(z,w)$, $g(z,w)$ such that 
they develop an $E_7$ singularity at $z=w=0$, and vary over 
$w$ according to Tate's algorithm \cite{Tate's} so that the singularity is relaxed 
to $A_4$.  
The result is \footnote{Note that this is not the most general 
equation for the curve. For instance $f(z,w)$ may contain a $z^5$ 
term but here it is set to zero for simplicity. Also the coefficient of 
the $z^3$ term needs not be 1.}
\beqa
f(z,w)&=&
-3 z^4+z^3+(a \epsilon-3 b^2) z^2+6 b \epsilon^2 z-3 \epsilon^4,
\label{f(z,w)}
\\
g(z,w)&=&2 z^6+\left(\frac{a^2}{12}+3 \epsilon^2+b\right) z^4
+(-2 b^3+a \epsilon b-\epsilon^2) z^3\n
&&+(6 b^2 \epsilon^2-a \epsilon^3) z^2-6 b \epsilon^4 z+2 \epsilon^6,
\label{g(z,w)}
\eeqa
where $a=a(w)$, $b=b(w)$ and $\epsilon=\epsilon(w)$ are 
smooth functions only of $w$ such that
\beqa
a(0)=b(0)=\epsilon(0)=0.
\eeqa
%
%
%

If one resorts to the general argument \cite{GaberdielZwiebach,DHIZ} on the 7-brane 
realization of the Kodaira singularities, it may easily be guessed what  
types of 7-branes are separating from the rest of coalesced branes. 
We can, however, directly see this by 
tracing the value of the $J$ function (see Appendix B). This technology was 
developed by Tani \cite{Tani}. For the purpose of illustration let us 
consider some special cases:
\begin{figure}[b]%
\centerline{
\includegraphics[width=1.0\textwidth]{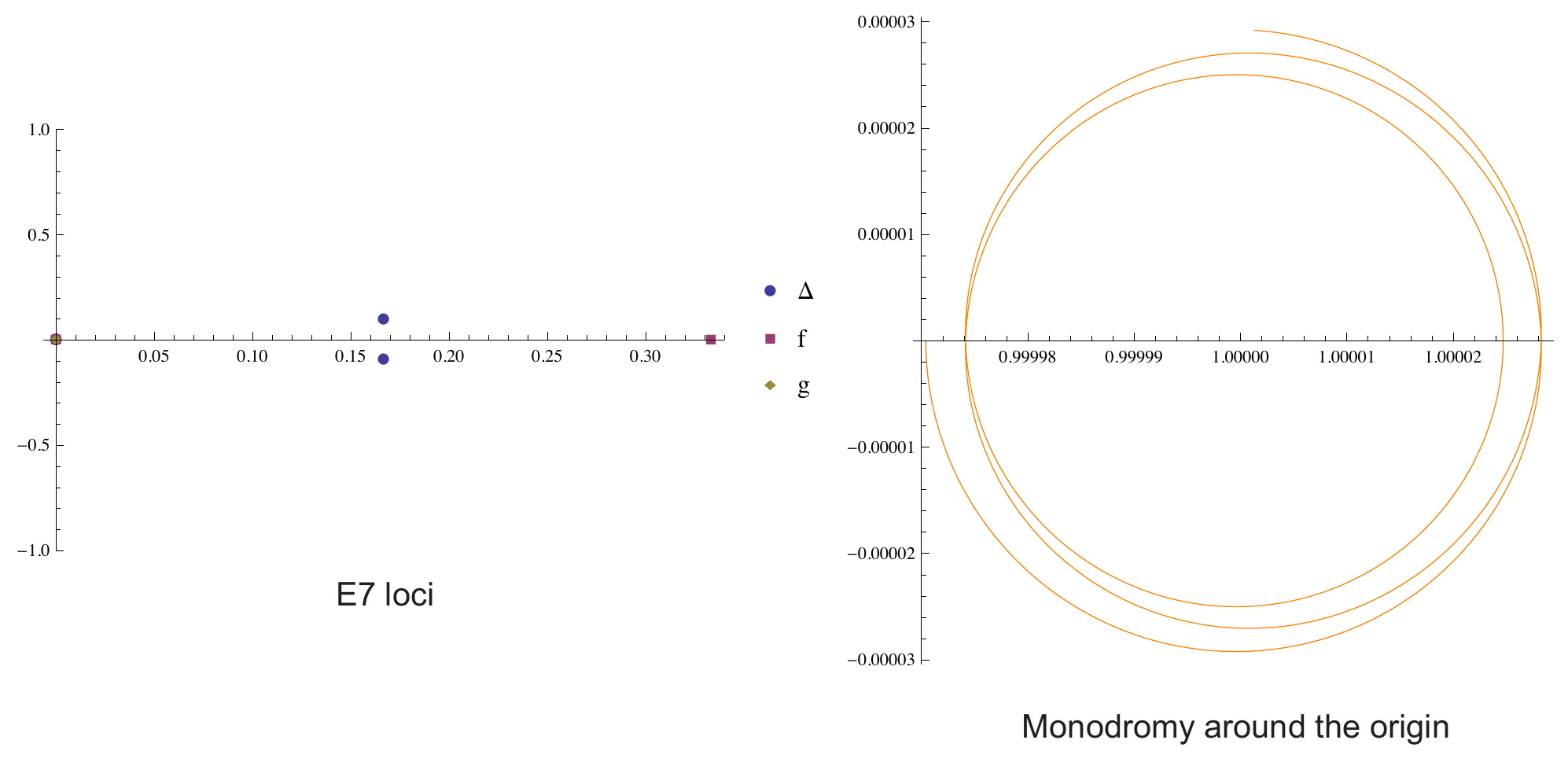}
}
\caption{\label{E7loci}  $E_7$ loci 
($a=b=\epsilon=0$). The monodromy is computed along a circle of 
radius$=0.01$ around the origin with the angle varying from $0$ 
to $2\pi - \frac\pi 6$ (and not to full $2\pi$, 
so that we may distinguish clockwise or anti-clockwise). This is anti-clockwise.}
\end{figure}
%

\paragraph{Case I : $a(w)=b(w)=\epsilon(w)=0$, unbroken $E_7$}
In this case (\ref{f(z,w)}) and (\ref{g(z,w)}) become simply
\beqa
f(z,w)&=&
-3 z^4+z^3,
\\
g(z,w)&=&2 z^6.
\eeqa
The discriminant is
\beqa
\Delta&=&4 z^9-36 z^{10}+108 z^{11},
\label{DeltaE7}
\eeqa
indicating that the curve has an $E_7$ singularity at $z=0$ for arbitrary $w$. 
This is a coalesced 7-brane configuration realizing unbroken 
$E_7$ gauge symmetry. The monodromy around these collapsed branes 
can be found by trancing the value of the $J$ function.
The  plot on the left of Figure \ref{E7loci} shows the locations of the roots of 
the discriminant $\Delta$ (\ref{DeltaE7}), and the plot on the right shows 
the contour of the value of the $J$ function when $z$ moves around the origin 
along the circle of radius $=0.002$.  
The latter shows that the value of $J$ moves three times around $J=1$ 
anti-clockwise, so the monodromy is 
\beqa
S^{-3}=S,
\eeqa
where
\beqa
S&\equiv&\left(
\begin{array}{cc}
 0 & -1     \\
 1 & 0
\end{array}
\right)
\eeqa
and $T$ (\ref{T}) are the fundamental generators of the $SL(2,\ZZ)$ group
 (see Appendix B). 
On the other hand,
\beqa
\AA^6\BB\CC^2&=&\AA^2
(\AA^4\BB\CC^2\AA^2)\AA^{-2},\\
\AA^4\BB\CC^2\AA^2&=&S,
\eeqa
which agrees with the monodromy read off from the behavior of the 
$J$ function.

\begin{figure}[h]%
\centerline{
\includegraphics[width=1.0\textwidth]{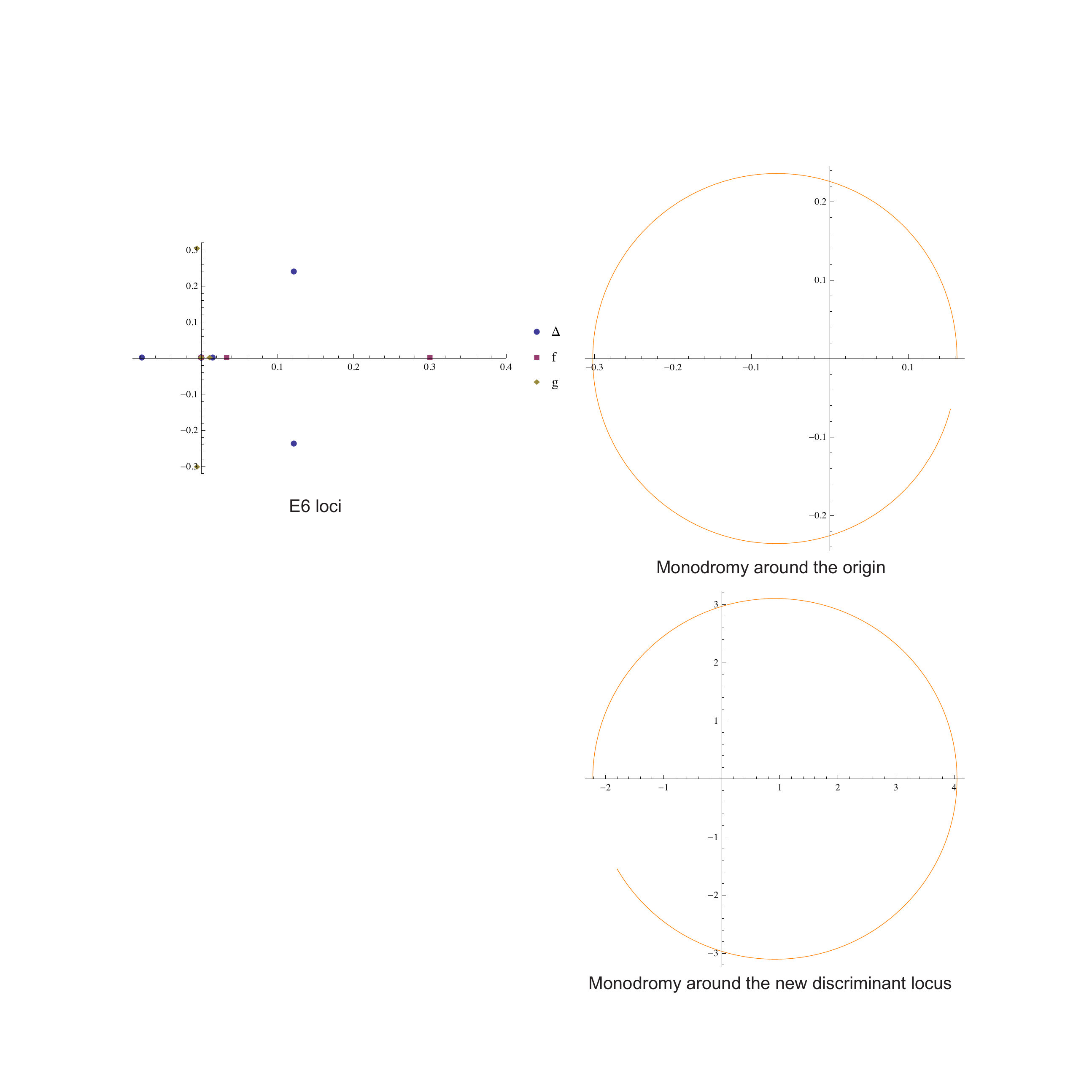}
}
\caption{\label{E6loci}  $E_6$ loci 
($a=1,b=\epsilon=0$). The radius of the circle around each point is $0.01$.
The range of the angle is from $0$ 
to $2\pi - \frac\pi 6$. }
\end{figure}

\paragraph{Case II : $b(w)=\epsilon(w)=0$ and $a(w)\neq 0$, $E_7/(E_6 \times U(1))$}
In this case the equations  (\ref{f(z,w)}), (\ref{g(z,w)}) read
\beqa
f(z,w)&=&
-3 z^4+z^3,
\\
g(z,w)&=&2 z^6+\frac{a(w)^2}{12} z^4,
\eeqa
which gives an $E_6$ singularity at $z=0$ for generic $w$ but 
it is enhanced to $E_7$ at $w=0$ since $a(0)=0$ as we assumed.
The discriminant in this case is
\beqa
\Delta&=&108 z^{11}+9 \left(a(w)^2-4\right) z^{10}+4 z^9+\frac{3 a(w)^4 z^8}{16},
\eeqa
whose roots are
\beqa
z=0 (\mbox{multiplicity eight}),
\left(\frac{1}{6}\pm\frac{i}{6 \sqrt{3}}\right)+\left(-\frac{1}{24}\pm\frac{i}{8
   \sqrt{3}}\right) a^2+O(a^4),
   -\frac{3 a^4}{64}+O(a^8).
\eeqa
The second to last complex conjugate pair is the loci of the 7-branes 
which were already separated from the coalesced branes in Case I,
while the last one is the position of the 7-brane bending into the 
transverse space. They are depicted in the left plot of Figure \ref{E6loci} for 
$a=1,b=\epsilon=0$. 

The monodromies around the origin 
and the locus 
of the separated brane are shown in the right plots. 
From these we can see that the former is $T^{-1}S$ as an 
$SL(2,\ZZ)$ conjugacy class, and the latter is $T$. Thus the separating 
brane is an \AA~brane, and the monodromy around the origin agrees 
with the fact that
\beqa
\AA^3\BB\CC^2\AA^2&=&T^{-1}S.
\eeqa
%

\begin{figure}[t]%
\includegraphics[width=1.0\textwidth]{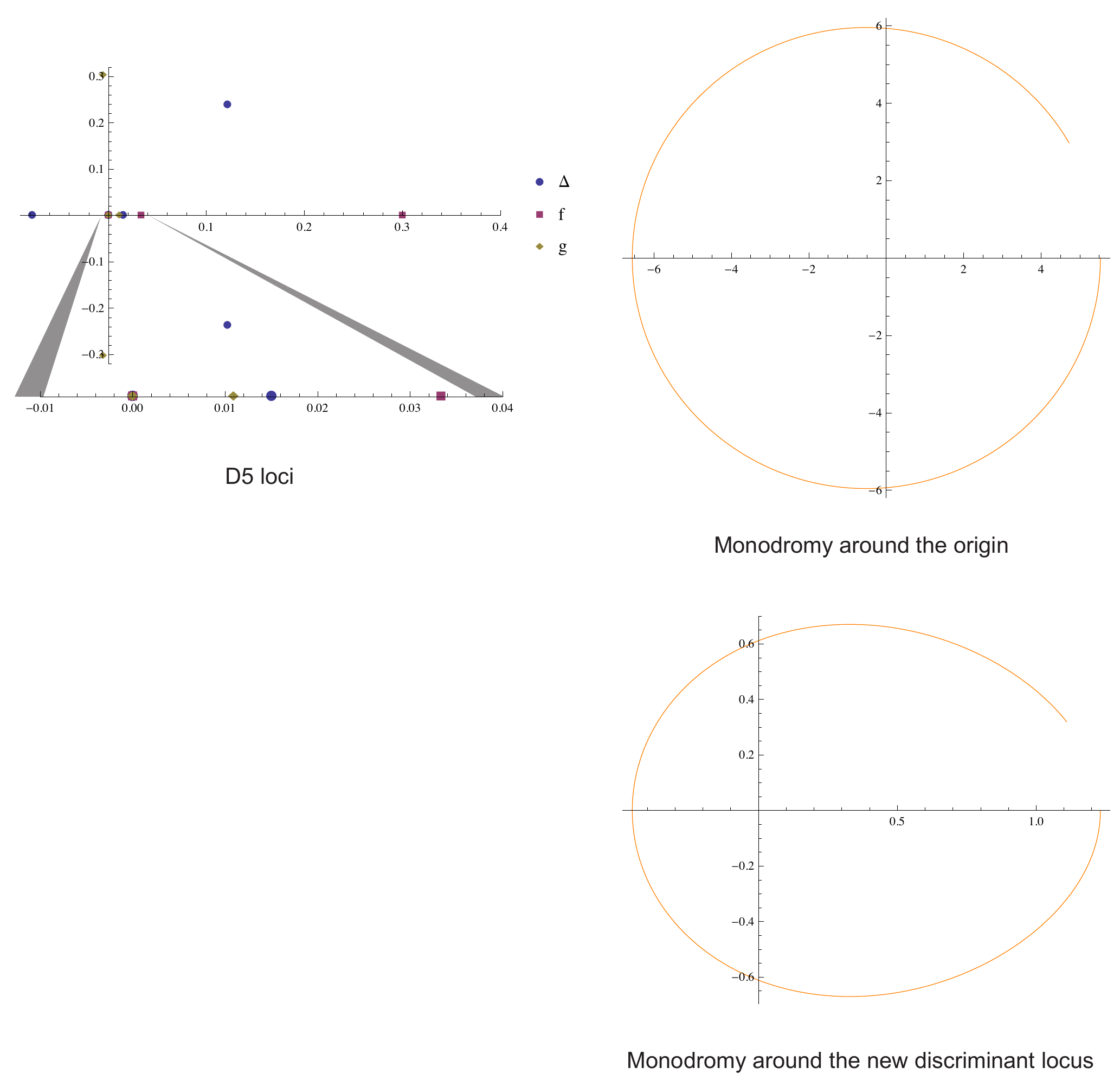}
\caption{\label{D5loci}  $D_5$ loci 
($a=1,b=0.1,\epsilon=0$). The radius is taken to be $0.002$ for 
both points. The angle for the origin is the same as before, and 
that for the new locus is taken from $-\pi$ to $\pi-\frac\pi 6$.}
\end{figure}

\paragraph{Case III : $\epsilon(w)=0$ and $a(w)b(w)\neq 0$, $E_7/(SO(10) \times U(1)^2)$}
If we also allow $b$ to take nonzero value, then we have 
\beqa
f(z,w)&=&
-3 z^4+z^3 -3 b^2 z^2,
\\
g(z,w)&=&2 z^6+\left(\frac{a^2}{12} +b\right) z^4
-2 b^3 z^3.
\eeqa
The discriminant takes the form
\beqa
\Delta&=&108 u^{11}+9 \left(a^2-36 b^2+12 b-4\right) u^{10}+\left(-216 b^3+216
   b^2+4\right) u^9\n
   &&+\left(\frac{3 a^4}{16}+\frac{9 b a^2}{2}-9 \left(36
   b^4+b^2\right)\right) u^8-9 a^2 b^3 u^7.
   \label{DeltaD5}
\eeqa
The new discriminant locus is
\beqa
\frac{48 b^3}{a^2}+\mbox{higher order in $b$},
\label{newD5locus}
\eeqa
which is shown in the plot on the left of Figure \ref{D5loci} as a blue 
circle near the origin ($a=1$, $b=0.1$). 

The monodromies around the origin and the locus (\ref{newD5locus}) 
are shown in the plots on the right. They tell us that both 
are $T$. However, we note that there is a locus of 
 $g(z)$ (the yellow diamond) 
between the two discriminant loci (blue circles), as shown in the enlarged figure. 
So if the reference point of the monodromy is taken near the origin, then 
to circle around the discriminant locus away from the origin  
one first needs to pass by the locus of $g(z)$ beforehand. Since the 
total monodromy around a locus of $g(z)$ is $S^{-2} =-1$, one gets $S^{-1}$ 
through a half rotation (anti-clockwise).
Thus the actual monodromy around the discriminant locus is 
the one obtained by the similarity transformation of the above:
$STS^{-1}$, which is equal to $T^{-1}S^{-1}T^{-1}$. Multiplying the monodromy 
around the origin $T$, 
we have
\beqa
T^{-1}S^{-1}T^{-1} \cdot T&=& T^{-1}S^{-1},
\eeqa
which is the same thing as $T^{-1}S$ in $PSL(2,\ZZ)$ and hence 
is consistent with Case II.
On the other hand,
\beqa
\AA^5 \BB \CC&\sim&\AA^{-1}(\AA^5 \BB \CC)\AA\n
&=&-T
\eeqa
which is equal to $T$ in $PSL(2,\ZZ)$, whereas
\beqa
\AA^{-1}\CC\AA&=&STS^{-1},
\eeqa
so the separating brane is indeed identified as the \CC~brane.

\paragraph{
Case IV : $a(w)b(w)\epsilon(w)\neq 0$, $E_7/(SU(5) \times U(1)^3)$}
The final case is when any of $a(w)$, $b(w)$ or $\epsilon(w)$ does not vanish. 
The functions $f$ and $g$ are given by (\ref{f(z,w)}) and (\ref{g(z,w)}), 
respectively, and the discriminant is
\begin{figure}[h]%
\includegraphics[width=1.0\textwidth]{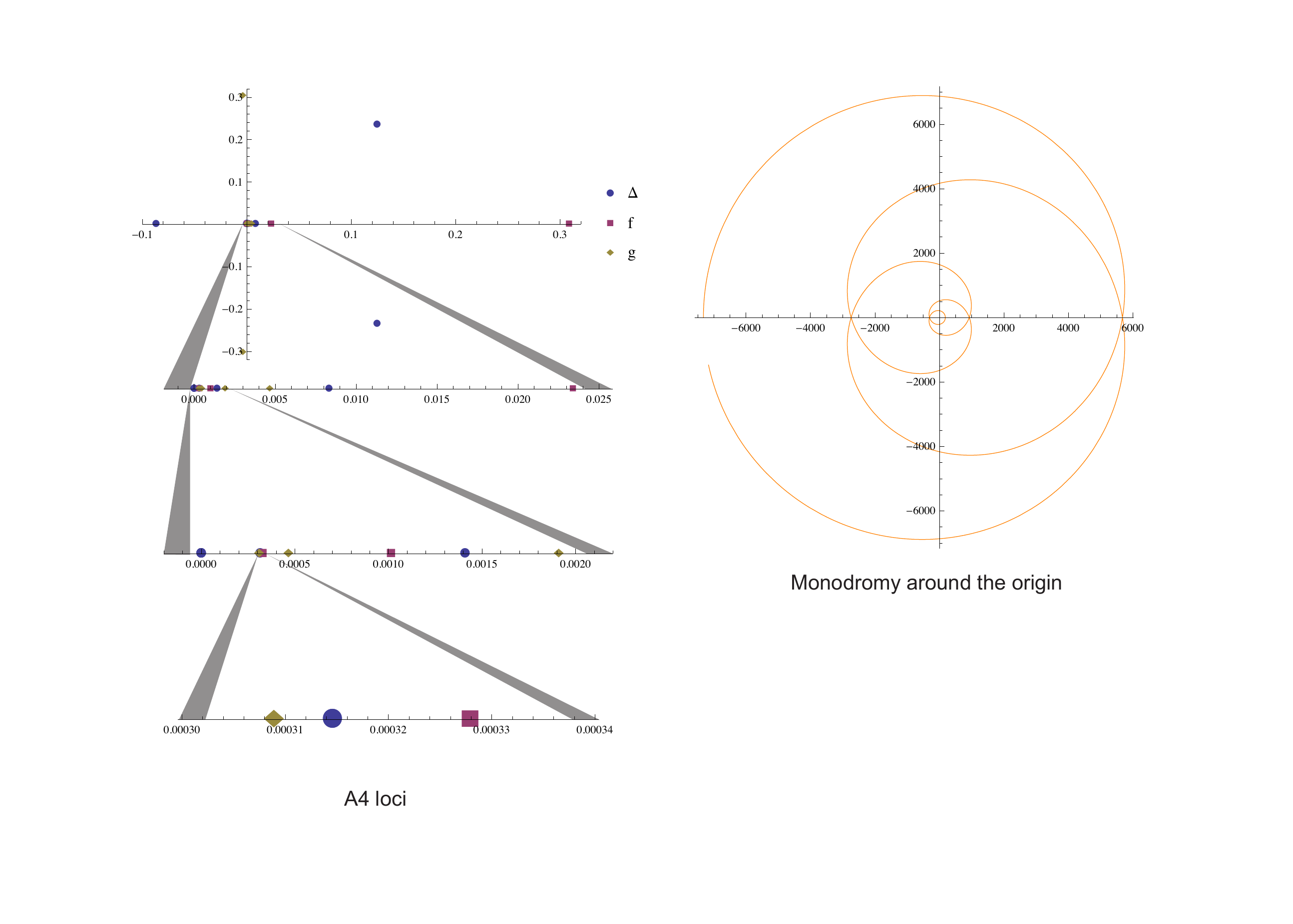}
\caption{\label{A4loci}  $A_4$ loci 
($a=1,b=0.1,\epsilon=0.007$). The radius is $0.0002$, and the angle 
is from $-\pi$ to $\pi-\frac \pi{60}$.}
\end{figure}
\beqa
\Delta&=&
108 u^{11}+9 \left(a^2+12 \epsilon  a+4 \left(-9 b^2+3 b+9 \epsilon
   ^2-1\right)\right) u^{10}\n
   &&+4 \left(-54 b^3+54 b^2+27 \epsilon  (a+6 \epsilon )
   b-27 \epsilon ^2-18 a \epsilon +1\right) u^9
   \n
   &&
   +\left(\frac{3
   a^4}{16}+\frac{9}{2} \left(b-5 \epsilon ^2\right) a^2-12 \epsilon  \left(-18
   b^2+9 \epsilon ^2-1\right) a\right.\n
   &&
   \left.~~~\rule{0ex}{3ex} -9 \left(36 b^4+\left(1-72 \epsilon ^2\right)
   b^2+30 \epsilon ^2 b+9 \epsilon ^4\right)\right) u^8
   \n
   &&+\frac{3}{2} \left(3 b
   \epsilon  a^3+\left(5 \epsilon ^2-6 b^3\right) a^2-12 b \epsilon  \left(15
   \epsilon ^2+b\right) a+12 \epsilon ^2 \left(54 b^3-36 \epsilon ^2 b+b+3
   \epsilon ^2\right)\right) u^7
   \n
   &&+\left(9 \left(-108 b^2+24 \epsilon ^2-1\right)
   \epsilon ^4-\frac{a^3 \epsilon ^3}{2}+18 a^2 b^2 \epsilon ^2+18 a \left(3
   \epsilon ^5+2 b \epsilon ^3\right)\right) u^6
   \n
   &&-9 \epsilon ^4 \left(b a^2+2
   \epsilon  a-36 b \epsilon ^2\right) u^5,
\eeqa 
which is of order $5$ as arranged for an $A_4$ singularity.
There appear two new nonzero discriminant loci; their positions are $\frac{\epsilon^2}b$ 
 for both to leading order in $\epsilon$, but different in their higher order terms.
The monodromy around the origin is $T^5$ as shown in Figure \ref{A4loci},
indicating the existence of five collapsed \AA ~branes. 
On the other hand, the monodromy around each discriminant locus 
is $T$ for both.  
Again, as shown in Figure \ref{A4loci}, there is a locus of $g$ between 
the origin and the closer locus, and there are 
two loci of $f$ and another locus of $g$ between the origin and the farther one.
Therefore we must do the corresponding similarity transformations. 
It can be shown that the monodromy is $S^{-1}$ for the former, and 
$T^2S^{-1}TST^{-2}$ 
for the latter. Thus the monodromy matrix for the closer locus is
\beqa
STS^{-1}=\AA^{-1}\CC \AA,
\eeqa
and that for the farther locus is
\beqa
(T^2S^{-1}TST^{-2})^{-1}T(T^2S^{-1}TST^{-2})&=&TS^{-1}T^{-3}\n
&=&\AA^{-1}(\CC^{-1} \BB\CC)\AA.
\eeqa
Changing their positions with each other in such a way that the farther 
brane passes through the branch cut of the closer brane, their monodromies 
become $\AA^{-1}\BB \AA$ and $\AA^{-1}\CC \AA$, so
they are a \BB~and a \CC~brane, respectively, as expected.

\section{Summary and discussion}
\label{Discussion}
Using Tani's argument to account for the chiral matter generation 
at extra zeroes, we have proposed a natural geometric mechanism 
for realizing the coset family structure in F-theory.
Tani's argument uses string junctions connecting the various gathering 
7-branes meeting at that point and is a direct generalization of \cite{BDL} for 
the intersecting D-brane systems. It offers a perfectly consistent picture of 
the chiral matter generation in six dimensions for the split-type singularities,
and we have pointed out that their relations to homogeneous K\"{a}hler 
manifolds are readily explained in this picture. 
Note that these rules are not just a kind of ``trick'' as has been known 
in the literature, but can be deduced 
by the concrete entities responsible for the symmetry enhancement: 
the string junctions.

In particular, we have 
proposed a local 7-brane system as shown in Figure \ref{spaghetti} which 
would yield the set of supermultiplets in six dimensions 
with exactly the same gauge 
quantum numbers as those in the three-generation 
$E_7/(SU(5)\times U(1)^3)$ coset family unification model. 
We have proved that for a given holomorphically varying type IIB 
scalar field configuration in six dimensions, there exists at least 
locally a K\"{a}hler metric such that a half of the supersymmetries   
are preserved.
We have further 
discussed how this local model is compactified to four dimensions and 
half of the spectrum is projected out on orbifolds, and also suggested 
how the anomalies of the original models can be canceled. The last point 
is still incomplete and we leave this issue to future work.

One of the nice features of our mechanism is that such a gathering 
brane system is a local one and could emerge independently of the 
every global detail of the ambient space. Meanwhile, 
given the experimental data, several authors have recently 
pointed out the possible existence of 
the ``desert'', the absence of new physics between the electro-weak
 and string scales \cite{IsoOrikasa,HamadaKawaiOda,Kawaidesert}. If this is true, 
and if string theory is really the theory beyond the Standard Model,
then it must have a mechanism to realize close to 
what we observe now already at the string scale. 
Our proposal fits with this requirement.

It is interesting to speculate 
how this configuration might come to exist: Suppose that there were, perhaps 
in the very early universe, some set of 7-branes, and assume some 
attractive force to be somehow generated between them. Then 
such 7-branes might have become closer and closer until 
they collide with each other.  This can happen only if they are 
a collapsable set of 7-branes which must be one of the types of 
the Kodaira classification. If these branes were the ones that could constitute 
the $E_7$ singularity, then just after they made a collision and at the last 
minute before they were completely separated, they would have 
looked like Figure \ref{spaghetti}.    
This story is of course just a speculation at this moment, but 
it would be an interesting scenario to study.

\acknowledgments
I wish to thank
T.~Higaki,
K.~Igi,
S.~Iso,
H.~Otsuka,
N.~Sakai,
Y.~Sakamura,
T.~Yukawa
and
Y.~Yasui
for 
useful discussions. My special thanks go to 
T.~Tani who explained his work.
I also thank M.~Yata for his advices in generating 
the figures.
The work of S.~M. is supported by Grant-in-Aid
for Scientific Research  
(A) \#22244430-0007 and 
 (C) \#25400285 from
The Ministry of Education, Culture, Sports, Science
and Technology of Japan.

\appendix
\section{
The explicit forms of $A_{(n, \bar n)}$ up to $n+\bar n\leq 3$}
\label{appendixA}
The result of solving the recursion relations (\ref{nBdelA}), (\ref{nAinvBBh}) is 
as follows:
\beqa
A_{(0,1)}&=& \frac{h_{(0,1)}}{h_{(0,0)}},\n
A_{(0,2)}&=& \frac{2
   h_{(0,2)}-B_{(0,1)}^{(0,1)}}{2 h_{(0,0)}},\n
A_{(1,1)}&=& \frac{B_{(0,1)}
    \overline{B_{(0,1)}}
   +h_{(1,1)}}{h_{(0,0)}},\n
A_{(0,3)}&=&\frac1{6 (h_{(0,0)})^3} 
\left(\rule{0ex}{3.5ex}
-2 B_{(0,2)}^{(0,1)} (h_{(0,0)})^2
-3
   B_{(0,1)}^{(0,1)} h_{(0,1)} h_{(0,0)}
   +3
   B_{(0,1)} h_{(0,1)}^{(0,1)} h_{(0,0)}\right.\n
   &&\left.
   -3
   B_{(0,1)} h_{(0,1)} h_{(0,0)}^{(0,1)}
   +6
   h_{(1,0)} (h_{(0,0)})^2
  \rule{0ex}{3.5ex} \right),\n
A_{(1,2)}&=&\frac1{ 2 (h_{(0,0)})^4}
\left(\rule{0ex}{3.5ex}
2 B_{(0,2)}(h_{(0,0)})^3
   \overline{B_{(0,1)}}
  \right.\n
   &&
   -(h_{(0,0)})^2
   \left(B_{(0,1)}^{(0,1)}h_{(1,0)}-2
   B_{(0,1)}
   h_{(1,0)}^{(0,1)}
   +h_{(0,1)}^{(1,1)}\right)
   \n
   &&+h_{(0,0)}
    \left(-2 B_{(0,1)}h_{(1,0)}
   h_{(0,0)}^{(0,1)}
   +h_{(0,1)}^{(1,0)}
   h_{(0,0)}^{(0,1)}+h_{(0,1)}^{(0,1)}
   h_{(0,0)}^{(1,0)}
   +h_{(0,1)}
   h_{(0,0)}^{(1,1)}\right)
   \n
   &&\left.+2 h_{(1,2)}
   (h_{(0,0)})^3
      -2 h_{(0,1)}h_{(0,0)}^{(0,1)}
   h_{(0,0)}^{(1,0)}\rule{0ex}{3.5ex}\right).\nonumber
\eeqa
$h_{(n,\bar n)}^{(p,q)}\equiv\partial_z^p \partial_{\bar z}^q h_{(n,\bar n)}$.
$B_{(0,\bar n)}$'s are arbitrary functions of $z$ and $\bar z$, and
$B_{(0,\bar n)}^{(p,q)}\equiv\partial_z^p \partial_{\bar z}^q B_{(0,\bar n)}$.
$A_{(\bar n,n)}$ is equal to the complex conjugate of $A_{(n,\bar n)}$. Once 
$A_{(n,\bar n)}$'s are determined, then so are
$B_{(n,\bar n)}$'s by the equation (\ref{nBdelA}).

\section{
$J$-function and monodromy}
\label{appendixB} 
\begin{figure}[b]%
\centering
\includegraphics[height=0.2\textheight]{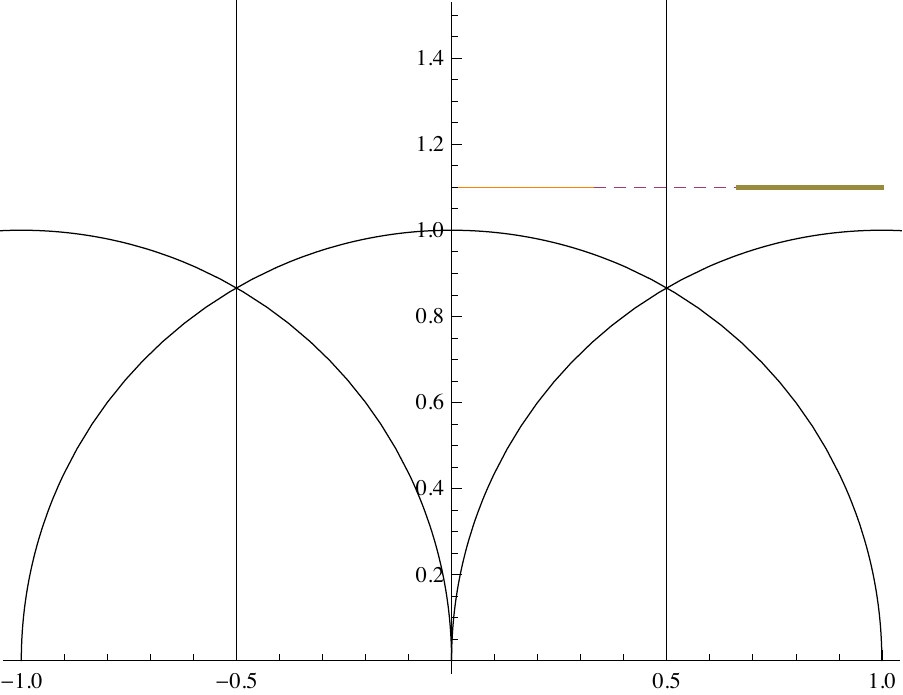}~~~~~
\includegraphics[height=0.2\textheight]{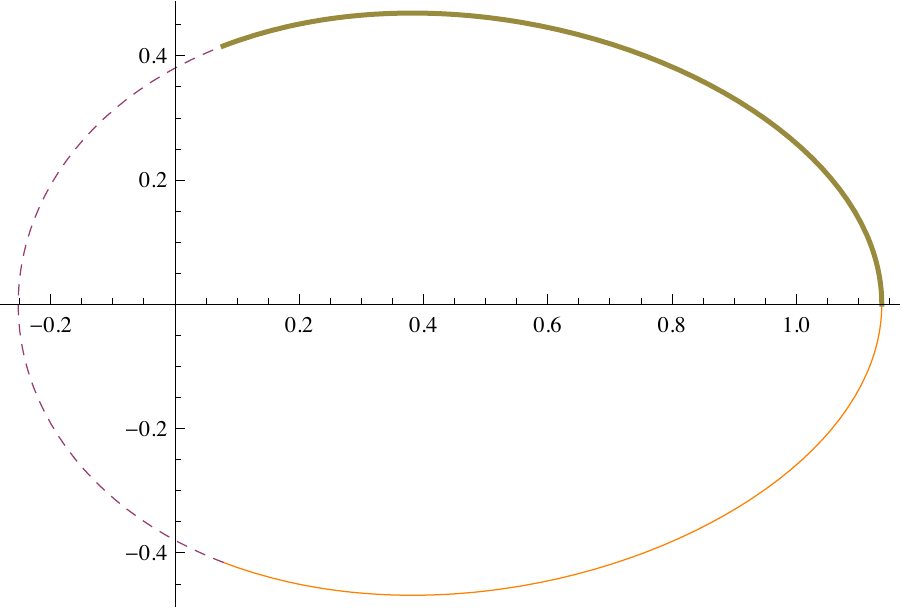}
\caption{\label{Trajectory}  A trajectory of $J(\tau)$ under the  $T$ 
transformation ($\tau_0=1.1i$). }
%
%
\includegraphics[height=0.2\textheight]{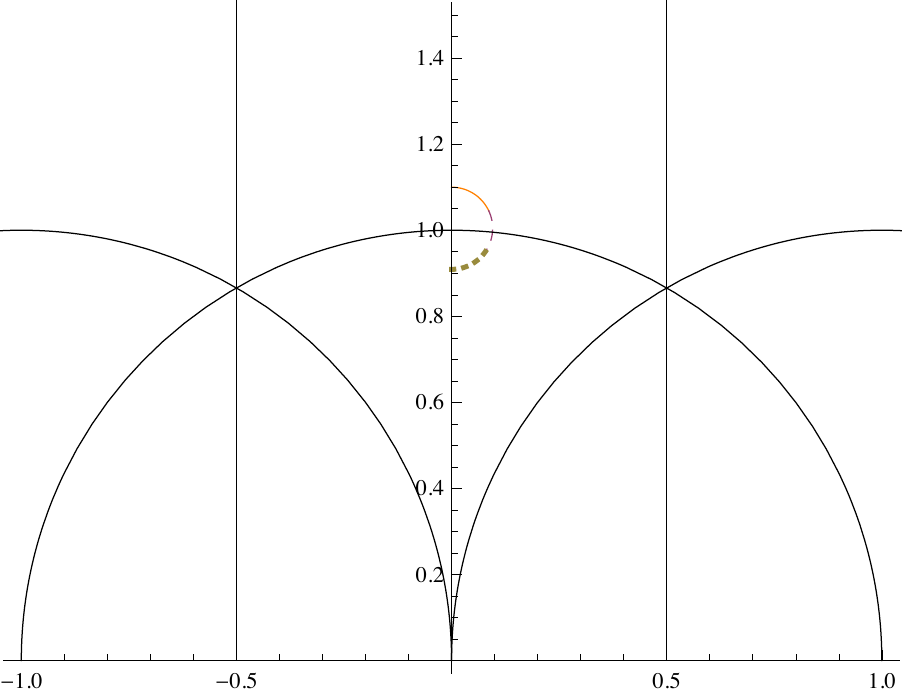}~~~~~
\includegraphics[height=0.2\textheight]{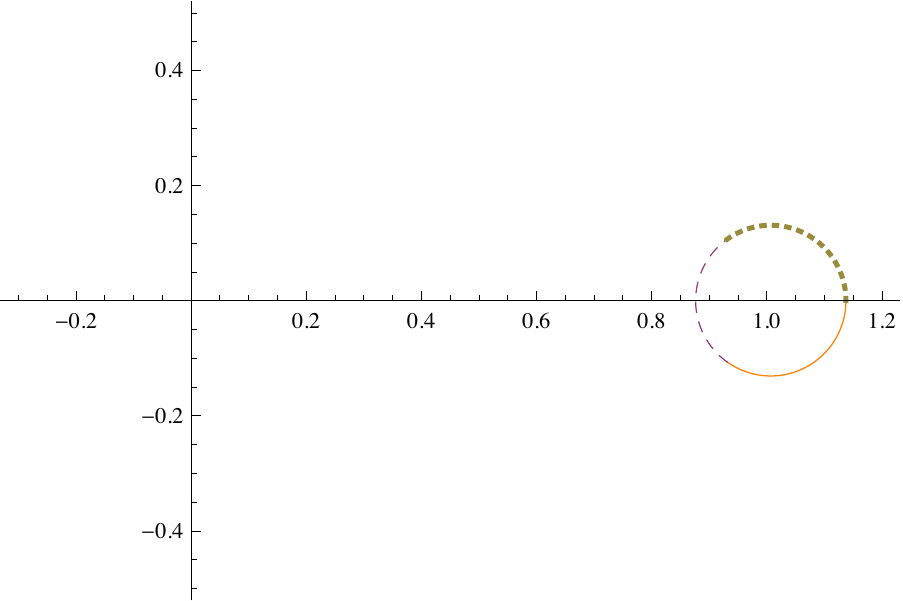}
\caption{\label{S}  A trajectory of $J(\tau)$ under the $S$ transformation
($\tau_0=1.1i$). }
\end{figure}

Klein's $J$-function is a modular invariant holomorphic function 
from the upper-half plane $\HH$ to the complex plane $\CCC$, and 
maps one-to-one 
the fundamental region of the modular group to the complex plane.
The definition in terms of theta functions is
\beqa
J(\tau)&=&\frac{(\vartheta_2(\tau)^8+\vartheta_3(\tau)^8+\vartheta_4(\tau)^8)^3}
{54\vartheta_2(\tau)^8\vartheta_3(\tau)^8\vartheta_4(\tau)^8},
\nonumber
\eeqa
so that
\beqa
J(e^{\frac{2\pi i}3})=0,~~~J(i)=1.
\nonumber
\eeqa

Suppose that $\tau\in\HH$ changes its value from some 
$\tau_0$ in the standard fundamental region to $\tau_0+1$,
which belongs to another fundamental region next to it. 
Then the trajectory of $J(\tau)$ circles clockwise 
around $1$ and $0$ (Figure  \ref{Trajectory}). On the other hand, if 
$\tau\in\HH$ changes from $\tau_0$ to $-\frac1{\tau_0}$,
then $J(\tau)$ only circles around $1$, counter-clockwise (Figure  \ref{S}).

Note that since $J(\tau)$ has a triple zero at $\tau=e^{\frac{2\pi i}3}$, 
$J(\tau)$ moves three times around $0$ when $\tau$ moves 
along a small circle around $e^{\frac{2\pi i}3}$
and back to the original fundamental region. Likewise 
$J(\tau)-1$ has a double zero at $\tau=i$, so $J(\tau)$ goes 
twice around $1$ when $\tau$ does 
once around $i$.

\end{document}